\documentclass{ieeeaccess}

\usepackage{cite}
\usepackage{amsmath,amssymb,amsfonts}
\usepackage{algorithmic}
\usepackage{graphicx}
\usepackage{textcomp}
\usepackage{array}
\usepackage{float}
\usepackage{tabularx}
\usepackage{ragged2e}
\usepackage[table]{xcolor}
\usepackage{pifont}
\newcommand{\cmark}{\ding{51}}
\newcommand{\xmark}{\ding{55}}

\renewcommand\arraystretch{1.15}

\usepackage{multirow}
\usepackage{booktabs}
\usepackage[table]{xcolor}
\usepackage{enumitem}
\usepackage{subcaption}
\usepackage{color}
\usepackage{listings}
\usepackage{float}
\usepackage{xcolor}
\usepackage{outlines}
\usepackage[normalem]{ulem}
\usepackage{cleveref}
\usepackage{soul}
\renewcommand{\hl}[1]{#1}
\usepackage{mdframed}

\usepackage{soul}
\usepackage{xcolor}
\usepackage[normalem]{ulem}

\definecolor{MyDarkGray}{RGB}{120,120,120}

\def\BibTeX{{\rm B\kern-.05em{\sc i\kern-.025em b}\kern-.08em
    T\kern-.1667em\lower.7ex\hbox{E}\kern-.125emX}}

\begin{document}

\history{Received xxxx xx, xxxx; accepted xxxx xx, xxxx; date of publication xxxx xx, xxxx; date of current version xxxx xx, xxxx.}
\doi{10.1109/ACCESS.XXXX.XXXXXXX}

\title{Natural Language Interfaces for Spatial and
Temporal Databases: A Comprehensive Overview of
Methods, Taxonomy, and Future Directions}

\author{\uppercase{Samya Acharja},
\uppercase{Kanchan Chowdhury}}
\address{Department of Computer Science, Marquette University, Milwaukee, Wisconsin, USA}

\corresp{Corresponding author: Kanchan Chowdhury (kanchan.chowdhury@marquette.edu).}

\begin{abstract}
The task of building a natural language interface to a database, known as NLIDB, has recently gained significant attention from both the database and Natural Language Processing (NLP) communities. With the proliferation of geospatial datasets driven by the rapid emergence of location-aware sensors, geospatial databases play a vital role in supporting geospatial applications. However, querying geospatial and temporal databases differs substantially from querying traditional relational databases due to the presence of geospatial topological operators and temporal operators. To bridge the gap between geospatial query languages and non-expert users, the geospatial research community has increasingly focused on developing NLIDBs for geospatial databases. Yet, existing research remains fragmented across systems, datasets, and methodological choices, making it difficult to clearly understand the landscape of existing methods, their strengths and weaknesses, and opportunities for future research. Existing surveys on NLIDBs focus on general-purpose database systems and do not treat geospatial and temporal databases as primary focus for analysis. To address this gap, this paper presents a comprehensive survey of studies on NLIDBs for geospatial and temporal databases. Specifically, we provide a detailed overview of datasets, evaluation metrics, and the taxonomy of the methods for geospatial and temporal NLIDBs, as well as a comparative analysis of the existing methods. Our survey reveals recurring trends in existing methods, substantial variation in datasets and evaluation practices, and several open challenges that continue to hinder progress in this area. Based on these findings, we identify promising directions for future research to advance natural language interfaces to geospatial and temporal databases.

\end{abstract}



\titlepgskip=-15pt
\maketitle

\section{Introduction}
\label{sec:intro}
We live in an era where conversational agents are gaining more prominence over static interfaces requiring specialized language expertise to query and retrieve data. Virtual agents with conversational ability such as Amazon Alexa and Google Assistant have become increasingly popular. While chatbots are used to give automatic response to user queries written in natural language, Amazon Alexa and Google Assistant allow users to communicate through natural language speech. In both the cases, these technologies need to understand the natural language query provided by the user and fetch the query answer from their data storage system. In order to use such technologies on systems where data is stored in databases, these automated question answering systems should be able to translate natural language questions to database queries, which emphasizes the need for \textit{Natural Language Interfaces to Databases}, shortly known as NLIDBs~\cite{zhong2017seq2sql, xu2017sqlnet, yu2018typesql, yu2018syntaxsqlnet}. NLIDBs allow database query-agnostic users who lack database expertise to issue queries to the database in natural language (NL), such as English. Given such an NL query, an NLIDB system automatically generates the corresponding database query~\cite{ml-4} and passes it to the database for execution. Because of the inherent complexity in translating NL questions to database queries, the task of building an efficient NLIDB is an open research problem. It should be noted that translating NL questions to database queries is different from predicting next database queries in a human-database interaction~\cite{next-sql-1, next-sql-2}.

The rapid growth of location-aware sensors has led to a sharp increase in spatial and spatiotemporal datasets, which supports a wide range of applications, such as mobility analytics and location-based services. To meet the resulting data management and query processing demands, many database systems now provide spatial and temporal extensions that introduce geometric data types, spatial indexing, and specialized operators for geospatial and temporal predicates. Representative examples include but are not limited to PostGIS~\cite{postgis} for spatial functionality in PostgreSQL, GeoSpark~\cite{geospark} for distributed geospatial analytics in SparkSQL, and MobilityDB~\cite{mobilitydb} for moving object functionality in PostgreSQL and PostGIS. However, writing database queries to retrieve and manage information is inherently challenging, and geospatial and temporal databases further increase this difficulty by introducing geospatial and temporal operators that often require domain expertise. That is why NLIDBs for geospatial and temporal databases~\cite{jiang2024chatgpt, wang2025gpt} have become an important research area, drawing interest from database, NLP, and geospatial research communities. Recent development of geospatial and temporal deep learning frameworks~\cite{pytorch-geotemporal, geotorch, geotorchai-1, geotorchai-2} also played a role in accelerating the research on spatial and temporal NLIDBs.


While the complex structure of database queries coupled with the inherent ambiguity of NL questions makes building an NLIDB system a challenging task, the spatial and temporal domains introduce extra challenges \cite{liu2024survey}\cite{fan2025rethinking}\cite{vossform}. Spatial and temporal datasets include many attributes that must be considered together, like geometry, topology, and relative distances, which makes the reasoning complex \cite{siampou2024poly2vec}. Besides, NLIDBs for spatial and temporal databases need to handle the semantic interpretation of spatial and temporal operators, such as \textit{ST\_Contains}, \textit{ST\_Distance}, etc. \cite{zhang2009natural, xu2023grammar}. 
Recent works have started adopting neural models \cite{zhou2024r3}, large language models \cite{liang2024nat}, and even multi-agent reasoning frameworks~\cite{liang2024nat}. However, these approaches vary widely, and the research community lacks a unified overview of the methods used by NLIDBs in this domain, the datasets they use, and their evaluated strategy \cite{chu2025geo2vec}.

\subsection{Limitations of Existing Surveys}
Although a substantial number of review papers already exist on NLIDBs, most of them primarily focus on traditional, non-spatial relational settings. For example, several surveys {\cite{nihalani2011natural,reshma2017review, zhang2024natural,liu2025systematic} review early rule-based and grammar-driven NLIDB systems, as well as neural approaches for translating NL questions into SQL, but without considering spatial or temporal semantics. Other works provide comparative analyses of NLIDB architectures and interaction paradigms \cite{affolter2019comparative}, or focus specifically on neural and deep learning models for NL to SQL (NL2SQL) tasks in standard relational databases \cite{iacob2020neural,wong2021survey}. More recent review papers discuss broader trends in NL interfaces, including advances in learning-based methods and system design \cite{das2023review,katsogiannis2023survey}, yet they do not explicitly address spatial predicates, spatiotemporal reasoning, or time-series querying.

Some surveys move closer to database-oriented perspectives by examining NL interfaces across different database paradigms \cite{liu2024survey,luo2025natural}, or by reflecting on future research directions for NLIDB systems \cite{fan2025rethinking,hong2025next}. However, even these works largely treat databases as abstract relational structures and do not engage with the unique challenges posed by spatial geometries, topological relations, temporal intervals, or trajectory data. As a result, spatial SQL languages, spatiotemporal query operators, and time-series–specific database constructs remain outside the scope of these surveys. Another set of related works, \cite{liu2025nli4db} \cite{liu2025survey} \cite{tang2025llm} \cite{singh2025survey}, broaden the discussion about NL interfaces for databases in general and provide useful information about system design and evaluation. \hl{Recent large language model (LLM)-focused surveys} {\cite{llm-survey-1, llm-survey-2, llm-survey-3, llm-survey-4, llm-survey-5, zhang2024large, huang2026role} \hl{further consolidate progress in LLM-based interfaces by reviewing modern architectures, learning-based methods, and evaluation practices across NLIDB systems.} However, the primary emphasis of these surveys remains on NLIDBs for general-purpose databases, and ST-NLIDBs are not treated as the primary target of analysis. \hl{Another category of LLM-based surveys} {\cite{jin2023large} \cite{ma2024survey} \hl{has explored large models for time-series and spatio-temporal data more broadly}. \hl{However, these studies focus on LLM architectures and pre-trained model paradigms for generic applications involving spatiotemporal and time-series data, rather than on natural-language interfaces for querying spatial or spatiotemporal databases.} \hl{A recent survey} {\cite{hashemi2026comprehensive} \hl{also examines agentic AI systems for spatio-temporal intelligence}. 
\hl{This survey focuses on agentic AI frameworks for spatio-temporal intelligence, emphasizing reasoning, tool use, and application domains rather than natural-language interfaces for querying spatial and temporal databases.}



Consequently, despite the richness of existing NLIDB survey literature, there is still a clear gap in comprehensive reviews that systematically focus on NLIDBs for spatial, spatiotemporal, and time-series databases, motivating the need for our study.

\subsection{Contributions}
\hl{This is the first comprehensive review that offers a detailed analysis of the existing research on NLIDBs for database systems in spatial and temporal domain, specifically geospatial, time-series, and spatiotemporal (moving object) databases.} In the rest of the paper, we use \textit{ST-NLIDBs} to refer to the NLIDB systems for geospatial, time-series, and spatiotemporal database systems. The review aims to provide a deep understanding of the main methods, datasets, and evaluation strategies used in the literature, while also identifying the major limitations and pointing toward possible future research directions. \hl{The key novelty of this study includes defining various ST-NLIDBs with examples, comparative analysis of benchmark datasets in terms of key characteristics and target NLIDBs, formalizing the evaluation metrics used for evaluating ST-NLIDBs, summarizing the common methodological pipeline for ST-NLIDBs, elaborating various classes of methodologies with key takeaway messages, advantages, and disadvantages, discussing the limitations of various methods and datasets, and proposing future research directions.} Our contributions can be summarized as follows.

\begin{itemize}[leftmargin=*]
    \item We outline various database systems in general-purpose and spatiotemporal domains and define corresponding NLIDBs with example queries (Section \ref{sec:background}).

    \item We summarize the datasets and evaluation metrics used for evaluating ST-NLIDBs. \hl{We provide a categorization of the benchmark datasets and list the key characteristics, advantages, and disadvantages of various datasets} (Section \ref{sec:data-eval-metric}).

    \item We summarize the common methodological pipelines among the NLIDBs in this domain (Section \ref{sec:overview-method}).

    \item \hl{We provide a taxonomy of the methods and discuss details under every class of methods with key takeaway messages, strengths, and limitations} (Section \ref{sec:method-taxonomy}).

    \item We identify key limitations of existing datasets and methods, and discuss future research directions (Section \ref{sec:discussions}).
\end{itemize}

\hl{We emphasize that this manuscript is a review article rather than an empirical study. Therefore, its purpose is not to present original experiments or benchmark results, but to analyze and critically discuss existing methods, datasets, evaluation metrics, strengths and limitations of current studies, and open research challenges.}

\section{Background}
\label{sec:background}
This section briefly outlines the inputs and outputs (I/Os) of NLIDBs for various database systems, in addition to classifying the database systems in terms of the presence of spatial and temporal features. At first, we define the I/Os of an NLIDB for a general-purpose relational database system. After that, we classify the spatial and temporal database systems and show sample I/Os for target NLIDBs.

\begin{figure*}[h!]
\begin{subfigure}{0.42\textwidth}
\centering
\includegraphics[width=\columnwidth]{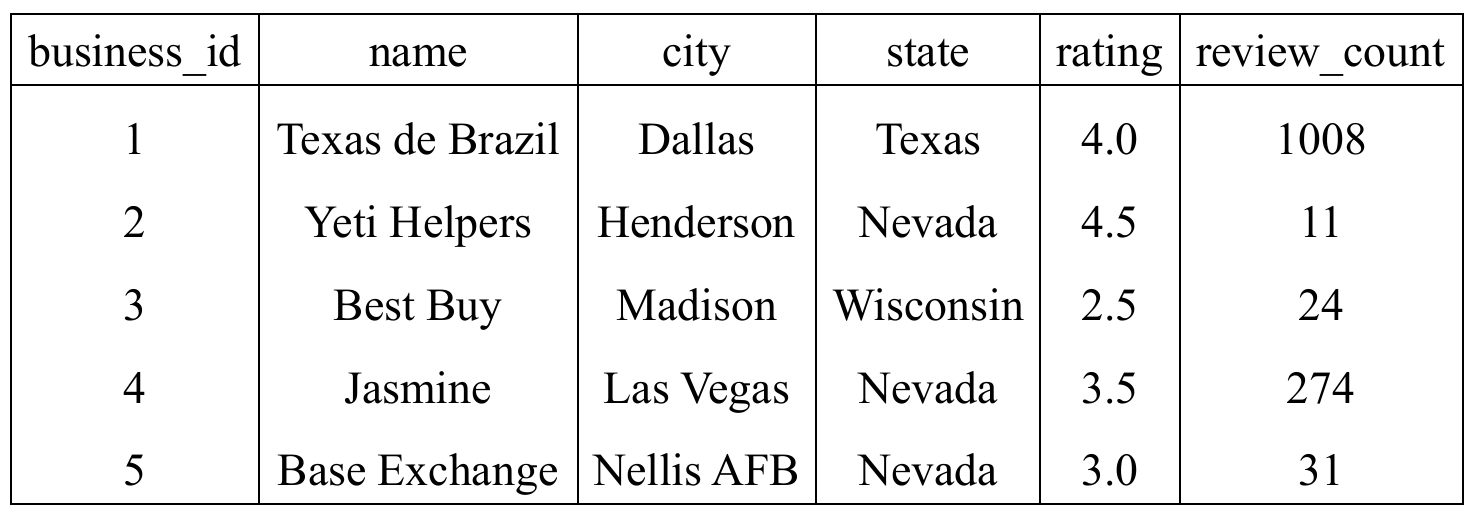}
\caption{Part of the business relation on Yelp Database}
\label{fig:business-relation}
\end{subfigure}%
\hspace{3mm}
\begin{subfigure}{0.55\textwidth}
\centering
\includegraphics[width=\columnwidth]{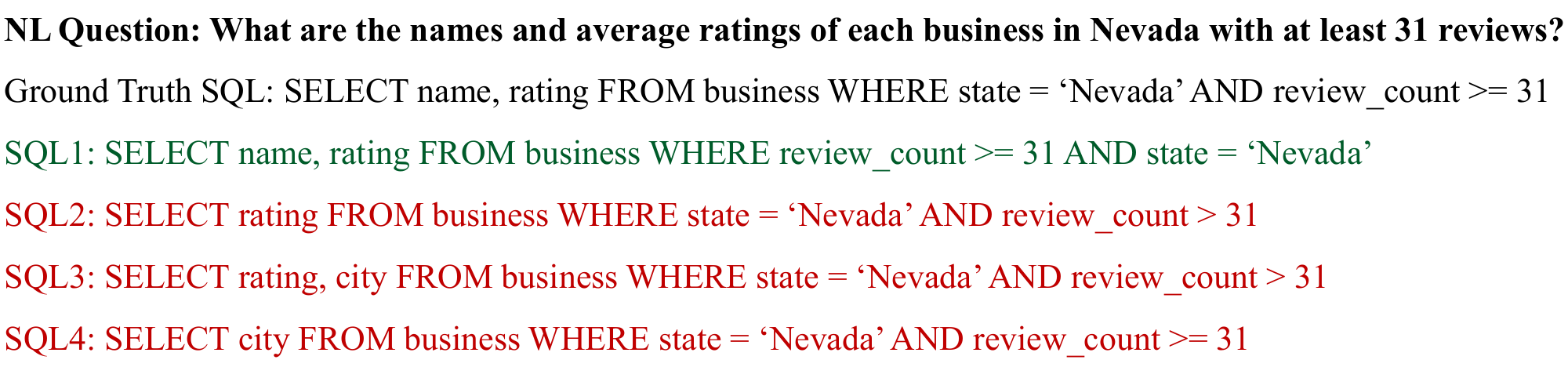}
\caption{An example NL question with ground truth SQL query and 4 SQL queries translated by NLIDBs}
\label{fig:example-question}
\end{subfigure}
\vspace{10pt}
\centering
\begin{subfigure}{.67\textwidth}
\includegraphics[width=\columnwidth]{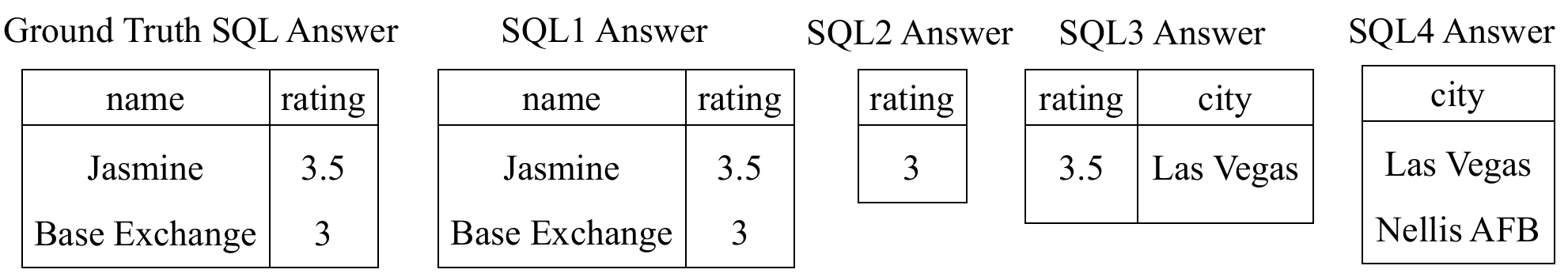}
\caption{Output Relations for SQL Queries}
\label{fig:output-relations}
\end{subfigure}
\vspace{-2mm}
\caption{A sample of the Business relation in Yelp, a sample natural language question with the ground truth SQL query, and 4 queries translated by various NL-2-SQL approaches as well as the output relations for such queries.}
\label{fig:example-relation}
\vspace{-3mm}
\end{figure*}

\subsection{General Purpose NLIDBs}
\label{subsec:gp-nlidb}
As stated earlier in Section \ref{sec:intro}, the task of an NLIDB is to translate an NL question into a database query, ultimately helping users who lack database expertise to retrieve information from databases. The target databases for these NLIDBs are usually relational databases which include but are not limited to PostgreSQL, Oracle, MySQL, and SQLite. Figure \ref{fig:example-relation} shows a sample dataset in the form of a relational table named \textit{business}, a sample input NL query and corresponding output database queries, and execution output of the ground truth database query.

The complex structure of SQL queries coupled with the inherent ambiguity of NL questions makes building an NLIDB system a challenging task. The difficulty in interpreting the correct semantic meaning of NL questions often results in the wrong synthesis of SQL queries. A small difference in the structure of a NL question might result in a huge difference in the semantic meaning of the question. Let us consider two NL queries for the relation \textit{business} in Figure \ref{fig:example-relation}: (1)~\textit{What is the number of businesses of all ratings in Nevada} and (2)~\textit{What is the total number of businesses in Nevada for each rating}. Although these two NL queries are almost similar, the formation of SQL query for the second one requires a GROUP BY operation whereas the SQL query corresponding to the first one does not have any GROUP BY operation. Besides this kind of syntactic and semantic ambiguities of NL questions, use of phrases and system specific terms makes the interpretation of NL questions much more difficult. The semantic meaning of a word varies based on the context in which it is used.

\subsection{ST-NLIDBs}
\label{subsec:st-nlidb}
\hl{By spatial and temporal databases, we refer to three categories of database systems: (i) spatial databases, (ii) time-series databases, and (iii) spatiotemporal/moving-object/trajectory databases.} ST-NLIDBs can be classified into three general classes based on their target database systems and support for spatial and temporal features: (i) spatial NLIDBs, (ii) time-series NLIDBs, and (iii) spatiotemporal NLIDBs.

\subsubsection{Spatial NLIDBs}
Most of the well-known relational databases support spatial extensions to support storage, querying, and management of geospatial datasets containing location coordinates. Under this category, we consider only databases or extensions of databases that have specialized support for location coordinates, but treat temporal columns similarly to generic relational databases. For example, PostGIS~\cite{postgis} extension provide native geospatial support for PostgreSQL database. Apache Sedona, formerly GeoSpark~\cite{geospark}, extends SparkSQL with geospatial support for distributed cluster computing in Apache Spark. SpatialHadop~\cite{spatialhadoop} adds native support for handling and querying massive geospatial datasets directly on Apache Hadoop's MapReduce architecture. SedonaDB~\cite{sedonadb} is a single-node analytical database engine that treats geospatial datasets as first-class citizen.

These spatial databases extend relational tables with a dedicated geometry or geography column that can store location coordinates as geometric objects, including points, line strings, and polygons, and more complex types such as multipolygons. This representation enables the database to model real-world entities, such as sensor locations, road networks, or administrative boundaries directly within database tables. To query and analyze these geospatial objects, spatial databases provide a rich library of spatial operators and functions, such as \textit{ST\_Contains} and \textit{ST\_Intersects} for topological relationships, and \textit{ST\_Distance} for distance-based filtering and ranking. Together with spatial indexing, these primitives allow efficient execution of common geospatial tasks such as point-in-polygon tests, proximity search, and spatial joins.

Spatial NLIDBs intend to translate NL questions to spatial database queries. This task is more challenging than general purpose NLIDBs because spatial NLIDBs need to handle the semantics of geometry objects and topological and distance relationships between geometry objects. Query Example 1 shows a spatial NLIDB query from nanjingtest\cite{liu2023nalspatial} dataset.

\vspace{5pt}
\noindent\fbox{%
  \begin{minipage}{\columnwidth}
    \textbf{Query Example 1:  Spatial NLIDB Query} \\
    \hrule \vspace{1mm} 
    \textbf{NL Query}: \textit{List the name of all POIs that are available in Jiangning District.}

\textbf{Executable Query (PostGIS)}:

\textcolor{blue}{SELECT} \textit{POI.name}

\textcolor{blue}{FROM} \textit{POI}

\textcolor{blue}{JOIN} \textit{district}

\textcolor{blue}{ON} \textit{ST\_Within(POI.geom, district.geom)}

\textcolor{blue}{WHERE} \textit{district.name= ‘Jiangning District’};
  \end{minipage}%
}

\subsubsection{Time-Series NLIDBs}
Time-series databases typically represent temporal data as records with an explicit timestamp column and often optimize physical storage and indexing around the time column. They commonly support time-aware operators and functions for windowed aggregation and trend analysis. Many database systems provide time-series functionality to support efficient querying and management of timestamped data. Under this category, we consider databases or extensions that treat time or timestamp as a first-class concept and provide specialized support for time-indexed data, while not necessarily providing native support for geospatial data. For example, TimescaleDB~\cite{TimescaleDB} extends PostgreSQL with hypertables and time-oriented query optimizations for scalable time-series analytics. InfluxDB~\cite{influxdb-applications} is a time-series database that supports time-centric querying for metrics and IoT data. These temporal analytics are supported through constructs such as \textit{time\_bucket}, continuous aggregates, and other built-in temporal operators.

The purpose of a time-series NLIDB is to translate NL questions to time-series database queries. Query Example 2 shows a time-series NLIDB query from InfluxDB’s NOAA water sample dataset.

\vspace{5pt}
\noindent\fbox{%
  \begin{minipage}{\columnwidth}
    \textbf{Query Example 2:  Time-Series NLIDB Query} \\
    \hrule \vspace{1mm} 
\textbf{NL Query}: \textit{How many water-level readings were recorded at the Coyote Creek station in 12-minute intervals between 00:00 and 00:30 UTC on Aug 18, 2019?}

\textbf{Executable Query (InfluxQL)}:

\textcolor{blue}{SELECT} \textit{COUNT(water\_level)}

\textcolor{blue}{FROM} \textit{h2o\_feet}

\textcolor{blue}{WHERE} \textit{location = ‘coyote\_creek’}

\textcolor{blue}{AND} \textit{time $\geq$ ‘2019-08-18T00:00:00Z’}

\textcolor{blue}{AND} \textit{time $\leq$ ‘2019-08-18T00:30:00Z’}

\textcolor{blue}{GROUP BY} \textit{time(12m)};
  \end{minipage}%
}

\subsubsection{Spatiotemporal NLIDBs}

In addition to purely spatial extensions, a growing class of systems provides specialized support for spatiotemporal or moving object data, where objects move or evolve over time and queries must jointly reason about \emph{where} and \emph{when}. Under this category, we consider databases and extensions that treat trajectories and time-indexed geometries as first-class entities, rather than simply pairing an ordinary timestamp column with a geometry column. 

\vspace{5pt}
\noindent\fbox{%
  \begin{minipage}{\columnwidth}
    \textbf{Query Example 3:  Spatiotemporal NLIDB Query} \\
    \hrule \vspace{1mm} 
\textbf{NL Query}: \textit{Find the 5 continuous nearest neighbors to train 100 between 6:00 and 19:00 o'clock.}

\textbf{Executable Query (SECONDO)}:

\textcolor{blue}{let} \textit{t100} = \textit{Trains} \textit{feed}
\textit{filter[.Id = 100]} \textit{extract[Trip]};

\textcolor{blue}{query} \textit{UTOrdered\_rtree}
\textit{UTOrdered} \textit{feed}
\textit{filter[(deftime(.UTrip) \textcolor{blue}{intersects}}
\textit{[const periods value (("2020-11-20-06:00" "2020-11-20-19:00" TRUE TRUE))])]}
\textit{\textcolor{blue}{nearest\_neighbor}[UTrip, t100, 5]} \textit{consume;}
  \end{minipage}%
}
\vspace{5pt}


\begin{table*}[t]
\centering
\small
\setlength{\tabcolsep}{6pt}
\renewcommand{\arraystretch}{1.18}
\caption{Major dataset families used in ST-NLIDB research: representative datasets, strengths and limitations}
\label{tab:dataset_strengths}

\begin{tabularx}{\textwidth}{p{3cm} p{4.8cm} X X}
\toprule
\textbf{Dataset Family} & \textbf{Representative Datasets} & \textbf{Strengths} & \textbf{Limitations} \\
\midrule

Benchmark datasets &
GeoQuery \cite{li2019spatialnli}, Restaurants \cite{li2019spatialnli}, GeoAnQu Corpus \cite{xu2023grammar}, GeoSQL-Bench \cite{hou2025geosql, houanl2sql}, TRUCE \cite{ito2024clasp}, SUSHI \cite{ito2024clasp}, TACO \cite{dohi2025retrieving}, NALMOBench \cite{wang2150nalmobench}, Text-to-PromQL Benchmark \cite{zhang2025promassistant}
&
Include predefined natural-language questions paired with executable queries, which makes it easier to evaluate whether a system correctly interprets spatial or temporal operators.
&
These datasets often do not reflect the ambiguity that appear in real-world GIS databases.
\\

Real-world GIS datasets &
Oracle Spatial Layers \cite{chintaphally2007extending}, 
Denton County GIS \cite{zhang2009natural}, 
TIGER/OSM/MODIS \cite{eldawy2015spatialhadoop}, 
berlintest \cite{liu2023nalsd, liu2025nalspatial}, 
nanjingtest \cite{liu2023nalsd, liu2025nalspatial}, 
chinawater \cite{liu2025nalspatial}, 
NYC GIS \cite{jiang2024chatgpt}, 
OSM Bavaria + SoilMap \cite{feng2025towards}, 
SpatialQueryQA \cite{khosravi2025questions}, 
KaggleDBQA \cite{khosravi2025questions, yu2025monkuu}, 
SpatiaLite \cite{wang2025gpt}, 
GeoQueryJP \cite{yu2025monkuu}, 
TourismQA \cite{yu2025spatial}
&
These datasets come from real GIS databases or OpenStreetMap layers, allowing NLIDB systems to be tested in realistic spatial environments.
&
These datasets usually do not include predefined natural-language query sets, so researchers often need to design their own evaluation queries, which makes results difficult to compare across different NLIDB systems.
\\

Synthetic / domain-specific datasets &
BerlinMOD trajectories \cite{duntgen2009berlinmod}, 
SECONDO trajectory demo \cite{wang2020nlmo}, 
MOQ + BT trains \cite{wang2021nalmo}, 
Extended MOQ + BT \cite{wang2023nalmo}
&
These datasets emphasize specific spatial or spatiotemporal query tasks, which helps researchers examine how NLIDB systems interpret trajectory relations, temporal predicates, and moving-object queries.
&
These datasets are artificially generated, often simplifying real geographic contexts, and therefore may not capture the data irregularities and interpretation challenges commonly found in real GIS databases.
\\

\bottomrule
\end{tabularx}
\end{table*}

Representative databases include MobilityDB~\cite{mobilitydb}, which extends PostgreSQL and PostGIS with native trajectory types and spatiotemporal operators for moving objects and SECONDO~\cite{secondo}, a research system that models moving regions and trajectories through dedicated spatiotemporal algebras. Spatiotemporal or moving object databases typically represent moving objects as sequences of time-stamped points, known as trajectories or as time-varying geometries, and they provide indexing schemes that couple spatial and temporal locality. This design enables efficient evaluation of queries that are difficult to express and optimize in standard spatial databases Accordingly, these systems support spatiotemporal predicates and operators for time slicing and temporal joins, distance-over-time and closest-approach queries, and trajectory similarity and pattern mining, in addition to traditional spatial predicates.

Spatiotemporal NLIDBs, also known as moving object NLIDBs, aim to translate natural-language questions into spatiotemporal database queries. Compared to spatial NLIDBs, this translation is more challenging because the interface must interpret temporal expressions and event constraints and map them to time-aware predicates, and correctly compose spatial and temporal semantics over trajectories. Note that spatiotemporal NLIDBs denote only one category ST-NLIDBs in our definition. Query Example 3 illustrates a representative spatiotemporal NLIDB query from BT Berlin Trains\cite{guting2005moving} dataset.

\section{Datasets and Evaluation Metrics}
\label{sec:data-eval-metric}
In this section, we provide a summary of the datasets and evaluation metrics used by NLIDBs for geospatial and temporal databases.

\subsection{Datasets Analysis}
Studies on ST-NLIDBs use a wide range of datasets that differ in spatial structure, temporal coverage, semantic richness, and their intended evaluation purposes. Although their methodological structures vary, the datasets used across the literature can still be grouped into three broad families: benchmark datasets, real-world geographic information system (GIS) datasets, and synthetic or domain-specific datasets. These categories reflect both the evolution of NL interfaces for spatial data and the different methodological objectives of the systems being evaluated.
\hl{Each of these dataset types plays a different role in system evaluation and comes with its own advantages and limitations. To give a quick overview before the detailed discussion of the datasets, Table} {\ref{tab:dataset_strengths} \hl{summarizes representative datasets in each family along with their main strengths and limitations.} To provide a clearer picture of the data used in existing work, Table~\ref{tab:data_extraction} summarizes the datasets employed across the included studies along with their key characteristics. 

As shown in the table, the datasets used by different systems are diverse. Many spatial NLIDB studies rely on real GIS datasets, including OpenStreetMap (OSM) layers and regional spatial databases, reflecting practical deployment scenarios. 
\hl{Time-series NLIDB approaches typically rely on benchmark-style datasets designed for retrieval or comparison tasks rather than direct database querying. Table} {\ref{tab:data_extraction} \hl{illustrates the categories and publication years of datasets used across various ST-NLIDBs.}

\begin{table*}[!t]
\centering
\scriptsize
\caption{\hl{Categories and publication years of datasets used across various ST-NLIDBs}}
\label{tab:data_extraction}
\begin{tabularx}{0.95\textwidth}{@{}
X
>{\centering\arraybackslash}p{0.06\textwidth}
*{5}{>{\centering\arraybackslash}p{0.085\textwidth}}
@{}}
\hline
\textbf{Dataset (Short Name)} &
\textbf{Year} &
\textbf{Spatial} &
\textbf{Temporal} &
\textbf{Benchmark} &
\textbf{RealGIS} &
\textbf{Synthetic} \\
\hline

Oracle Spatial Layers \cite{chintaphally2007extending} & \hl{2007} &
\cmark & \xmark & \xmark & \cmark & \xmark \\

Denton County GIS \cite{zhang2009natural} & \hl{2009} &
\cmark & \xmark & \xmark & \cmark & \xmark \\

BerlinMOD Trajectories \cite{duntgen2009berlinmod} & \hl{2009} &
\cmark & \cmark & \cmark & \xmark & \cmark \\

TIGER / OSM / MODIS \cite{eldawy2015spatialhadoop} & \hl{2015} &
\cmark & \xmark & \xmark & \cmark & \xmark \\

GeoQuery / Restaurants \cite{li2019spatialnli} & \hl{2019} &
\cmark & \xmark & \cmark & \xmark & \xmark \\

SECONDO Traj (Demo) \cite{wang2020nlmo} & \hl{2020} &
\cmark & \cmark & \xmark & \xmark & \cmark \\

MOQ + BT Trains \cite{wang2021nalmo} & \hl{2021} &
\cmark & \cmark & \cmark & \cmark & \cmark \\

GeoAnQu Corpus \cite{xu2023grammar} & \hl{2023} &
\cmark & \xmark & \cmark & \xmark & \xmark \\

Extended MOQ + BT \cite{wang2023nalmo} & \hl{2023} &
\cmark & \cmark & \cmark & \cmark & \cmark \\

berlintest / nanjingtest \cite{liu2023nalsd} & \hl{2023} &
\cmark & \xmark & \xmark & \cmark & \xmark \\

NYC GIS + Socioeconomic DB \cite{jiang2024chatgpt} & \hl{2024} &
\cmark & \cmark & \xmark & \cmark & \xmark \\

TRUCE / SUSHI TS \cite{ito2024clasp} & \hl{2024} &
\xmark & \cmark & \cmark & \xmark & \cmark \\

OSM Bavaria + SoilMap \cite{feng2025towards} & \hl{2025} &
\cmark & \xmark & \xmark & \cmark & \xmark \\

TACO Time-Series Differences \cite{dohi2025retrieving} & \hl{2025} &
\xmark & \cmark & \cmark & \xmark & \cmark \\

GeoSQL-Bench \cite{hou2025geosql, houanl2sql} & \hl{2025} &
\cmark & \xmark & \cmark & \xmark & \cmark \\

SpatialQueryQA + KaggleDBQA \cite{khosravi2025questions} & \hl{2025} &
\cmark & \xmark & \cmark & \cmark & \xmark \\

SpatiaLite (Ada / Edu / Tour / Traffic) \cite{wang2025gpt} & \hl{2025} &
\cmark & \xmark & \cmark & \cmark & \xmark \\

GeoQueryJP / KaggleDBQA \cite{yu2025monkuu} & \hl{2025} &
\cmark & \xmark & \cmark & \cmark & \xmark \\

berlintest / nanjingtest / chinawater \cite{liu2025nalspatial} & \hl{2025} &
\cmark & \xmark & \xmark & \cmark & \xmark \\

TourismQA + MapQA \cite{yu2025spatial} & \hl{2025} &
\cmark & \xmark & \xmark & \cmark & \xmark \\

NALMOBench \cite{wang2150nalmobench} & \hl{2025} &
\cmark & \cmark & \cmark & \cmark & \cmark \\

Text-to-PromQL Benchmark \cite{zhang2025promassistant} & \hl{2025} &
\xmark & \cmark & \cmark & \xmark & \xmark \\

\hline
\end{tabularx}
\end{table*}

Table~\ref{tab:dataset_features} provides a detailed overview of the datasets used across the studies included in this review, along with their key characteristics. The table is designed to highlight how different types of data support NL interfaces for spatial, spatiotemporal, and time-series databases, and to make explicit the assumptions each study makes about data availability, query formulation, and evaluation setup.

The column labeled \textit{\#NLQs} reports the number of NL queries associated with a dataset when such queries are explicitly defined. The value \emph{N/A} is used in cases where a dataset does not provide a fixed or explicitly enumerated set of NL queries. 

The \textit{Query Language} column describes the formal language or representation to which NL queries are translated. Here, \textit{SQL} refers to standard relational SQL, while \textit{Spatial SQL}\cite{ramsey2005introduction} denotes SQL dialects extended with spatial data types and operators, such as those supported by PostGIS. \textit{Geo-Analytical QL}\cite{xu2023grammar} refers to higher-level geo-analytical query formalisms that express spatial analysis workflows or transformations rather than direct database queries, as used in geo-analytical question answering systems. For spatiotemporal systems, languages such as SECONDO represent executable algebras that natively support moving objects and temporal operators, whereas time-series benchmarks rely on specialized query or difference operators designed for signal retrieval and comparison. In addition to SQL-based and algebraic query languages, some datasets rely on domain-specific query languages tailored to particular data ecosystems. In particular, \textit{OverpassQL} is used by datasets such as OverpassNL~\cite{Text-to-OverpassQL} to query OSM data~\cite{bennett2010openstreetmap} via the Overpass API\cite{olbricht2011overpass}. OverpassQL is a declarative, read-only query language that supports spatial constraints and semantic tag-based filtering over OSM entities (nodes, ways, and relations), and is distinct from Spatial SQL dialects used in database systems such as PostGIS. We therefore list OverpassQL separately in the \emph{Query Language} column to reflect its OSM-specific data model and execution environment.
The \textit{NLQ Origin} column indicates how the NL queries were created, distinguishing between human-authored queries, synthetically generated queries, or mixed settings. The \textit{Split} column specifies whether an explicit train–test or evaluation split is defined, which is common for benchmark-style datasets but often absent in operational or demonstration-oriented systems. Finally, the \textit{Public} column indicates whether the dataset is publicly available.
By organizing datasets into operational GIS databases, NL query corpora, time-series benchmarks, executable NL to query benchmarks, and NL to action datasets, Table~\ref{tab:dataset_features} shows that existing NLIDB research relies on a wide range of data sources with very different assumptions. This diversity helps explain differences in system design and evaluation practices across studies, but also highlights the lack of standardized datasets that NL querying in a unified manner.

\begin{table*}[!t]
\centering
\scriptsize
\caption{Key characteristics of datasets across all included studies}
\label{tab:dataset_features}
\begin{tabularx}{\textwidth}{@{}
p{0.17\textwidth}
p{0.23\textwidth}
>{\centering\arraybackslash}p{0.07\textwidth}
>{\centering\arraybackslash}p{0.14\textwidth}
>{\centering\arraybackslash}p{0.14\textwidth}
>{\centering\arraybackslash}p{0.06\textwidth}
>{\centering\arraybackslash}p{0.06\textwidth}
@{}}
\toprule
\textbf{Dataset / Data Source} &
\textbf{Paper(s) Using Dataset} &
\textbf{\#NLQs} &
\textbf{Query Language} &
\textbf{NLQ Origin} &
\textbf{Split} &
\textbf{Public} \\
\midrule

\multicolumn{7}{l}{\textit{Operational GIS Databases (Execution-Centric)}} \\
\midrule

Oracle Spatial Layers\cite{chintaphally2007extending} &
\cite{chintaphally2007extending} &
N/A & Spatial SQL & None & \xmark & \cmark \\

Denton County GIS\cite{abugovoracle} &
\cite{zhang2009natural} &
30 & SQL & Human & \xmark & \cmark \\

BerlinMOD Trajectories\cite{duntgen2009berlinmod} &
\cite{duntgen2009berlinmod} &
N/A & SECONDO & None & \xmark & \cmark \\

SECONDO Demo Traj\cite{guting2005moving} &
\cite{wang2020nlmo} &
N/A & SECONDO & None & \xmark & \cmark \\

BT Berlin Trains\cite{guting2005moving} &
\cite{wang2021nalmo}, \cite{wang2023nalmo} &
N/A & SECONDO & None & \xmark & \cmark \\

berlintest\cite{yaramanci2002aquifer} / nanjingtest\cite{liu2023nalspatial} / chinawater\cite{liu2025nalspatial} &
\cite{liu2023nalsd, liu2023nalspatial, liu2025nalspatial} &
N/A & SECONDO & None & \xmark & \xmark \\

NYC GIS + Socioecon\cite{xu2020extracting} &
\cite{jiang2024chatgpt} &
$\sim$200 & SQL & Mixed & \xmark & \cmark \\

\midrule
\multicolumn{7}{l}{\textit{Natural Language Query Corpora}} \\
\midrule

GeoQuery\cite{thompson1997learning} &
\cite{li2019spatialnli} &
880 & $\lambda$-calculus / SQL & Human & \cmark & \cmark \\

Restaurants\cite{liang2013learning} &
\cite{li2019spatialnli} &
$\sim$300 & $\lambda$-calculus & Human & \cmark & \cmark \\

GeoAnQu Corpus\cite{xu2023grammar} &
\cite{xu2023grammar} &
309 & Geo-Analytical QL & Human & \xmark & \cmark \\

MOQ (2021)\cite{wang2021nalmo} &
\cite{wang2021nalmo} &
240 & SECONDO & Synthetic & \xmark & \cmark \\

Extended MOQ (2023)\cite{wang2023nalmo} &
\cite{wang2023nalmo} &
280 & SECONDO & Synthetic & \xmark & \cmark \\

GeoQueryJP\cite{yu2025monkuu} &
\cite{yu2025monkuu} &
$\sim$500 & Spatial SQL & Human & \cmark & \cmark \\

\midrule
\multicolumn{7}{l}{\textit{Time-Series Language–Signal Benchmarks}} \\
\midrule

TRUCE\cite{truce2021} &
\cite{ito2024clasp} &
$\sim$300 & TS Query Ops & Human & \cmark & \cmark \\

SUSHI\cite{sushi2022} &
\cite{ito2024clasp} &
$\sim$300 & TS Query Ops & Human & \cmark & \cmark \\

TACO TS Differences\cite{taco-dataset} &
\cite{dohi2025retrieving} &
$\sim$100 & TS-Diff Ops & Synthetic & \cmark & \cmark \\

\midrule
\multicolumn{7}{l}{\textit{Executable NL-to-Query Benchmarks}} \\
\midrule

GeoSQL-Bench\cite{houanl2sql} &
\cite{houanl2sql}, \cite{hou2025geosql} &
14,178 & Spatial SQL & Synthetic & \cmark & \cmark \\

SpatialQueryQA\cite{khosravi2025questions} &
\cite{khosravi2025questions} &
90 & SQL & Human & \cmark & \cmark \\

KaggleDBQA\cite{KaggleDBQA} &
\cite{khosravi2025questions}, \cite{yu2025monkuu} &
$\sim$2,500 & SQL & Human & \cmark & \cmark \\

SpatiaLite (Ada/Edu/Tourism/Traffic)\cite{furieri2011spatialite} &
\cite{wang2025gpt} &
$\sim$50 & SQL & Mixed & \xmark & \cmark \\

OverpassNL\cite{Text-to-OverpassQL} &
\cite{Text-to-OverpassQL} &
8,352 & OverpassQL & Mixed & \cmark & \cmark \\

NALMOBench~\cite{wang2150nalmobench} &
\cite{wang2150nalmobench} &
600 & SECONDO & Mixed & \cmark & \cmark \\

\midrule
\multicolumn{7}{l}{\textit{Natural Language to Action / API Datasets}} \\
\midrule

IoD-520 (Drone API)\cite{drones9060444} &
\cite{drones9060444} &
520 & REST API & Mixed & \cmark & \xmark \\

\bottomrule
\end{tabularx}
\end{table*}

\begin{table*}[t]
\centering
\small
\setlength{\tabcolsep}{6pt}
\renewcommand{\arraystretch}{1.18}
\caption{\hl{Significance of various evaluation metrics across spatial, spatiotemporal, and time-series NLIDB systems}}
\label{tab:evaluation_metrics_twocol}

\rowcolors{2}{gray!6}{white}
\begin{tabular}{p{2.5cm} p{4.6cm} p{4.6cm} p{4.6cm}}
\toprule
\rowcolor{gray!12}
\textbf{Evaluation Dimension} &
\textbf{Spatial NLIDB} &
\textbf{Spatiotemporal NLIDB} &
\textbf{Time-Series NLIDB} \\
\midrule

Execution-Based accuracy &
Correct execution of spatial queries &
Correct execution of time-aware queries &
Not applicable \\

Interpretation / Translation Accuracy &
Correct mapping of NL queries to spatial intent &
Correct identification of spatiotemporal intent &
Correct alignment of text with signal semantics \\

Precision-Oriented accuracy &
Precision of spatial results &
Precision of trajectory or temporal outputs &
Precision of retrieved time-series segments \\

Structural / Syntactic accuracy &
Optional SQL structure matching &
Rarely used &
Not applicable \\

Retrieval Quality (Ranking-Based) &
Rare &
Occasional &
Primary (e.g., mAP) \\

Efficiency and Scalability &
Runtime and performance &
Trajectory-length and time-range efficiency &
Embedding and retrieval latency \\

Qualitative / Human Evaluation &
Manual spatial inspection &
Manual trajectory inspection &
Manual semantic inspection \\

Benchmark-Specific Composite Metrics &
Rare &
Occasional &
Occasional \\

\bottomrule
\end{tabular}
\end{table*}

Benchmark datasets form a substantial part of the empirical foundation of NLIDB research. These datasets are designed to be clean, structured, and suitable for evaluating different methods. For example, GeoSQL-Bench and GeoSQL-Eval \cite{houanl2sql, hou2025geosql} generate thousands of synthetic GeoSQL tasks targeted at operator-level reasoning, covering distance predicates, topological relations, buffering, spatial joins, and multi-table queries. Semantic parsing corpora such as GeoQuery~\cite{thompson1997learning} and its variants used in earlier work like SpatialNLI \cite{li2019spatialnli} provide canonical NL questions paired with executable logical forms. More recent benchmarks extend this idea beyond static geometry. TRUCE~\cite{truce2021} and SUSHI \cite{ito2024clasp, dohi2025retrieving} incorporate NL descriptions of temporal signals and time-series differences, expanding evaluation into the temporal domain. The key strength of these benchmark datasets is their ability to isolate reasoning behavior. If a system succeeds or fails on a benchmark, it usually points to whether the issue lies in operator selection, schema grounding, or some part of the logical interpretation. But at the same time, these datasets rarely include the kind of noise, irregularity, or semantic ambiguity that shows up in practical GIS deployments. Because of that, strong benchmark performance does not always translate well to real-world robustness.

Real-world GIS datasets fall into a different kind of evaluation setting, where the goal is more about ecological validity than having everything neatly controlled. Systems like NALSD \cite{liu2023nalsd}, NALSpatial \cite{liu2025nalspatial}, and NALMO \cite{wang2021nalmo, wang2023nalmo} usually rely on operational SECONDO spatial databases such as \textit{berlintest}~\cite{liu2023nalsd}, \textit{nanjingtest}~\cite{liu2023nalsd}, or \textit{chinawater}~\cite{liu2025nalspatial}. 

More recent LLM based systems, for example ChatGPT for PostGIS \cite{jiang2024chatgpt} and multi agent frameworks \cite{khosravi2025questions, feng2025towards}, tend to use municipal GIS layers, OSM derived data, and a few socio economic datasets as well. These real world spatial databases often come with irregular geometries, inconsistent naming, missing attributes, and multi layer schemas that reflect historical or operational quirks rather than any ideal relational design.

Because of all that, evaluating spatial and temporal NLIDB systems on real GIS data brings out challenges that benchmark datasets do not reveal, like schema ambiguity, geographic noise, entity matching issues, and even spatial variability from place to place. In that sense, real world GIS datasets reveal the actual robustness of spatial NLIDB systems, especially the ones that depend on LLM prompting or schema-aware reasoning.



Synthetic and domain specific datasets occupy a third methodological structure, where they target the evaluation of particular spatial or temporal reasoning cases that are hard to isolate with real data. Moving object datasets created through the SECONDO trajectory model, used in NLMO and NALMO \cite{wang2020nlmo, wang2021nalmo, wang2023nalmo}, make it possible to test dynamic predicates such as \textit{overlaps}, \textit{during}, \textit{enters}, \textit{leaves}, and \textit{trajectory similarity}. The BerlinMOD vehicle movement benchmark \cite{duntgen2009berlinmod} has become a common dataset for testing spatiotemporal query processing in this area.

More recent systems also generate synthetic thematic databases that are tailored for LLM evaluation \cite{houanl2sql, hou2025geosql}, which allows fine grained testing of edge case geometries, complex joins, and operator accuracy. Time series datasets such as TRUCE~\cite{truce2021} and SUSHI \cite{ito2024clasp, dohi2025retrieving} provide NL supervision for temporal patterns and temporal reasoning.

Synthetic datasets offer a high degree of experimental control and help researchers test very specific reasoning abilities. At the same time, they can miss the contextual and semantic richness of real-world environments, and because of that they may oversimplify the challenges that are usually expected in operational GIS systems.

Putting everything together, these dataset families reveal both conceptual progress and practical fragmentation in the research of ST-NLIDBs. Still, no single dataset family really captures all the dimensions that are needed for a full and complete evaluation. Benchmark datasets usually underrepresent ambiguity and geographic noise. Synthetic datasets often simplify the semantic context more than they should. 
Therefore, it becomes hard to judge performance across general settings. The field would benefit a lot from having unified benchmark suites that mix controlled operator diversity with real-world geographic irregularity and natural user questions, allowing a more solid and holistic evaluation of spatial and spatiotemporal NL interfaces. 

\vspace{5pt}
\noindent\fbox{%
  \begin{minipage}{\columnwidth}
   
\textbf{Key Takeaway}: \textit{\hl{Current ST-NLIDB research relies on three main dataset types: benchmark datasets, real-world GIS databases, and synthetic datasets, each serving different evaluation goals. While benchmarks enable controlled testing and real GIS data reveal practical challenges, no single dataset type captures the full complexity of natural language interaction with spatial databases, highlighting the need for more unified and realistic evaluation benchmarks.}}

  \end{minipage}%
}

\subsection{Evaluation Metrics}

Evaluation of ST-NLIDBs is inherently more heterogeneous than in standard NL to SQL (NL2SQL) settings. A key reason is that multiple formal queries can express the same user intent, particularly when spatial predicates (distance buffers, containment relations, spatial joins) or temporal constructs (time windows, intervals, trajectory segments) are involved. As a result, many studies focus less on reproducing an exact reference query and more on whether a system produces a semantically correct outcome.

Table~\ref{tab:evaluation_metrics_twocol} summarizes the main evaluation dimensions reported across the included studies and illustrates how their interpretation varies across modalities. These dimensions capture evaluation goals rather than specific numeric formulas, and together they reflect the diverse ways NLIDB systems are assessed.

\paragraph{Execution-Based accuracy.}
The most common evaluation dimension in ST-NLIDB systems is \textit{execution-based accuracy}. Under this criterion, a system is considered correct if the generated query executes successfully and returns the expected objects, regions, or trajectories. Importantly, syntactic equivalence with a reference query is not required. This outcome-oriented evaluation is particularly suitable for geospatial queries, where semantically equivalent alternatives are frequent and strict string matching would be misleading. Execution-based accuracy is generally not applicable to time-series interfaces that do not produce executable database queries.

\paragraph{Interpretation / Translation Accuracy.}
Another important evaluation dimension is \textit{interpretation or translation accuracy}, which focuses on whether a natural-language query can be correctly translated into a meaningful internal representation. Rather than checking for an exact match with a reference query, this metric evaluates whether the system produces a semantically complete and executable formulation that reflects the user’s intent.

In executable-language, this notion is commonly captured through the concept of \emph{translatability} \cite{wang2150nalmobench}. A natural-language query is considered correctly interpreted if the system can generate an executable query that preserves the required entities and logical structure, even if the output obtained from execution on database differs from that of the gold query. Interpretation accuracy is therefore reported as the proportion of input queries that can be successfully translated into valid executable queries:
\begin{equation}
\text{TA} = \frac{|EQ|}{|N|}
\end{equation}
where $EQ$ denotes the set of executable queries generated by the system and $N$ denotes the total number of natural-language queries.

This interpretation-oriented metric is particularly appropriate for ST-NLIDB systems, where multiple executable formulations may express the same user intent and strict exact-match evaluation would be overly restrictive.

\paragraph{Precision-Oriented accuracy.}
In addition to interpretation accuracy, several systems report \textit{precision-oriented accuracy},
which is also referred to as \emph{translation precision} \cite{wang2150nalmobench}. A translated query is considered precise if it not only executes successfully, but also produces results that are equivalent to those of the corresponding gold query. In this setting, equivalence means same execution results. Precision-oriented accuracy is typically computed as:

\begin{equation}
\text{TP} = \frac{1}{N} \sum_{n=1}^{N} \big( \mathbb{I}(S_n, G_n) \cdot \mathbb{I}(R(S_n), R(G_n)) \big),
\end{equation}
where $\mathbb{I}(A, B)$ is an indicator function defined as
\[
\mathbb{I}(A, B) =
\begin{cases}
1, & \text{if } A = B, \\
0, & \text{otherwise}.
\end{cases}
\]
Here, $S_n$ and $G_n$ denote the generated and gold executable queries for the $n$-th example, respectively, and $R(\cdot)$ denotes the corresponding execution result set.

\paragraph{Retrieval Quality (Ranking-Based).}
For time-series NLIDB systems and retrieval-oriented interfaces, \emph{retrieval quality} becomes a central evaluation dimension. Rather than generating executable queries, these systems retrieve relevant sequences or segments based on learned representations, and are therefore evaluated using ranking-based metrics. Common examples include mean average precision (mAP), Recall@k, or related information-retrieval measures:
\[
\text{mAP} = \frac{1}{Q} \sum_{q=1}^{Q} \text{Average Precision}(q),
\]
where \(Q\) denotes the number of queries. Such metrics are primary in time-series NLIDBs, occasional in spatiotemporal systems, and rare in purely spatial NLIDBs.

\paragraph{Efficiency and Scalability.}
Beyond accuracy, many studies report \emph{efficiency and scalability} metrics. These include runtime, translation latency, query execution time, and scaling behavior with respect to data size or query complexity. In spatial databases, performance is often influenced by indexing and spatial joins, while trajectory length and temporal range impact the spatiotemporal systems. Time-series systems are mainly affected by embedding computation and retrieval latency.

\paragraph{Qualitative / Human Evaluation.}
A number of systems rely on \emph{qualitative or human evaluation}, where outputs are manually inspected for semantic accuracy, spatial reasoning quality, or practical usefulness. This evaluation dimension is common in early-stage systems or exploratory studies and typically complements quantitative metrics rather than replacing them.

\paragraph{Benchmark-Specific Composite Metrics.}
Finally, a smaller set of works introduces \textit{benchmark-specific composite metrics} that combine multiple evaluation aspects, such as syntactic validity, execution success, robustness, or partial accuracy, into a single framework-defined score. While these metrics facilitate comparison within a specific benchmark, they are often not directly comparable across different evaluation setups.

\vspace{5pt}
\noindent\fbox{%
  \begin{minipage}{\columnwidth}
   
\textbf{Key Takeaway}: \textit{\hl{Although metric names and reporting styles vary widely across studies, the underlying evaluation objectives tend to align with a small number of dimensions. The lack of consistent terminology across modalities complicates cross-study comparison and highlights the need for better reporting practices and more widely adopted benchmarks for ST-NLIDB systems.}}

  \end{minipage}%
}




\section{Overview of Methods}
\label{sec:overview-method}
In this section, we do not jump into how various ST-NLIDBs handle spatial and temporal semantics of their datasets methodologically, which will be detailed in Section \ref{sec:method-taxonomy}. Instead, we initially present a structured summary of ST-NLIDBs, where we focus on the types of models that have been proposed, the datasets they used, the database systems they targeted, and the range of query types they addressed. Then, we discuss various methodological pipelines which are common in various ST-NLIDBs. 

\begin{table*}[!t]
\centering
\scriptsize
\caption{Summary of included studies with evaluation metrics and NLIDB category}
\label{tab:included_studies_normalized}
\begin{tabularx}{0.98\textwidth}{@{}
p{0.08\textwidth}
p{0.05\textwidth}
p{0.23\textwidth}
p{0.18\textwidth}
p{0.22\textwidth}
p{0.12\textwidth}@{}}
\hline
\textbf{Paper} &
\textbf{Year} &
\textbf{Model Type} &
\textbf{DBMS} &
\textbf{Evaluation Metric (Normalized)} &
\textbf{NLIDB Category} \\
\hline

\cite{wang2000fuzzy} &
2000 &
Fuzzy grammar + possibility theory-based semantic parsing &
Relational GIS (SQL preprocessor) &
Qualitative / Human Evaluation &
Spatial NLIDB \\

\cite{chintaphally2007extending} &
2007 &
Menu-based NLI + rule-based grammar &
Oracle Spatial &
Qualitative / Human Evaluation &
Spatial NLIDB \\

\cite{zhang2009natural} &
2009 &
Rule-based NLP + semantic parsing &
PostGIS &
Execution-Based Accuracy &
Spatial NLIDB \\

\cite{duntgen2009berlinmod} &
2009 &
Data generation and benchmark framework &
SECONDO &
Efficiency and Scalability &
spatiotemporal NLIDB \\

\cite{eldawy2015spatialhadoop} &
2015 &
MapReduce spatial framework (non-NLI) &
HDFS / SpatialHadoop &
Efficiency and Scalability &
Spatial NLIDB (Context) \\

\cite{li2019spatialnli} &
2019 &
Seq2Seq GRU + spatial comprehension model &
Custom geospatial DB &
Interpretation / Translation Accuracy &
Spatial NLIDB \\

\cite{wang2020nlmo} &
2020 &
Rule-based NLP + structured language generation &
SECONDO &
Execution-Based Accuracy &
spatiotemporal NLIDB \\

\cite{wang2021nalmo} &
2021 &
Hybrid NLP + LSTM classifier + rules &
SECONDO &
Interpretation / Translation Accuracy; Precision-Oriented Correctness &
spatiotemporal NLIDB \\

\cite{xu2023grammar} &
2023 &
Rule-based CFG, concept transformations &
No DBMS (abstract workflows) &
Qualitative / Human Evaluation &
Spatial NLIDB \\

\cite{wang2023nalmo} &
2023 &
Hybrid semantic parsing + LSTM &
SECONDO &
Interpretation / Translation Accuracy &
spatiotemporal NLIDB \\

\cite{liu2023nalsd} &
2023 &
NLP + LSTM classifier + rule templates &
SECONDO &
Interpretation / Translation Accuracy; Efficiency and Scalability &
spatiotemporal NLIDB \\

\cite{ito2024clasp} &
2024 &
Contrastive learning (signal + text embeddings) &
N/A (time-series learning) &
Retrieval Quality (Ranking-Based); Qualitative / Human Evaluation &
Time-Series NLIDB \\

\cite{jiang2024chatgpt} &
2024 &
LLM prompting for spatial SQL &
PostGIS (NYC datasets) &
Execution-Based Accuracy &
Spatial NLIDB \\

\cite{khosravi2025questions} &
2025 &
Multi-agent spatial text-to-SQL &
PostGIS &
Qualitative / Human Evaluation &
Spatial NLIDB \\

\cite{dohi2025retrieving} &
2025 &
Contrastive retrieval model &
N/A (TACO-based TS pairs) &
Retrieval Quality (Ranking-Based) &
Time-Series NLIDB \\

\cite{feng2025towards} &
2025 &
Multi-agent LLM + semantic search &
Vector DB + OSM layers &
Qualitative  / Human Evaluation &
Spatial NLIDB \\

\cite{houanl2sql} &
2025 &
LLM evaluation framework &
PostGIS (synthetic thematic DBs) &
Benchmark-Specific Composite Metrics; Structural / Syntactic Correctness &
Spatial NLIDB \\

\cite{wang2025gpt} &
2025 &
GPT-based text-to-SQL with schema prompts &
SpatiaLite &
Execution-Based Accuracy &
Spatial NLIDB \\

\cite{yu2025monkuu} &
2025 &
LLM + dynamic schema mapping &
PostGIS + GeoQueryJP &
Translation Accuracy &
Spatial NLIDB \\

\cite{liu2025nalspatial} &
2025 &
NLP + LSTM type classifier + rule-based SLM &
SECONDO &
Execution-Based Accuracy; Interpretation / Translation Accuracy &
Spatial NLIDB \\

\cite{redd2025queries} &
2025 &
Multi-agent LLM pipeline (ReAct-style reasoning + tool use) &
Relational SQL DBs (location-based check-in datasets) &
Execution-Based Accuracy; Qualitative / Human Evaluation &
Spatial NLIDB \\

\cite{wang2150nalmobench} &
2025 &
Text-to-EXE benchmark + evaluation (LLMs + baselines) &
SECONDO (executable language for MOD) &
Benchmark-Specific Composite Metrics; Execution-Based Accuracy &
spatiotemporal NLIDB \\

\cite{zhang2025promassistant} &
2025 &
LLM + knowledge retrieval for text-to-PromQL &
Prometheus / PromQL &
Interpretation / Translation Accuracy; Precision-Oriented Correctness &
Time-Series NLIDB \\

\hline
\end{tabularx}
\end{table*}

\subsection{Summary of Studies}
Among the studies in spatiotemporal NLIDBs, we observed a wide range of system designs, database platforms, and methodological choices. Earlier studies mostly relied on rule-based parsing or handcrafted grammars, while more recent studies have started to adopt neural models, multi-agent LLM pipelines, and larger evaluation setups.

The studies also differ in the types of geospatial or spatiotemporal databases they target, such as PostGIS, Oracle Spatial, SECONDO, SpatiaLite, and a mix of synthetic or benchmark datasets. They also vary in the kinds of queries they handle. For example, some studies focus only on spatial operators like within, intersects, or nearest neighbor, while others cover spatiotemporal queries, or time-series comparisons. A smaller group of studies deal with multi-step spatial reasoning and schema-aware SQL generation using LLM-based agents. Evaluation approaches range from basic qualitative examples to execution accuracy, semantic correctness, and ranking-based metrics. Table~\ref{tab:included_studies_normalized} provides a summary of the included studies by listing the publication year, model type, database system, supported query types, and the evaluation metrics reported in each study.

\subsection{Common Methodological Pipeline}
As stated earlier in this section, this subsection aims at discussing the common pipelines in various methodologies, instead of diving deeper into how spatial and temporal semantics are handled by these NLIDBs, which will be discussed later in Section \ref{sec:method-taxonomy}. Across research on spatial and spatiotemporal NLIDBs, a fairly consistent methodological pattern appears. Whether a system is designed for static GIS data, moving objects, or time-series data, most approaches rely on a multi-stage translation process. In this process, an NL query is gradually transformed into an executable structured form, such as SQL, GeoSQL, SECONDO algebra, or another domain-specific query language.

\hl{Figure}~{\ref{fig:method_pipeline} \hl{shows the unified methodological pipeline commonly used in ST-NLIDBs, illustrating how natural language questions are progressively transformed into executable database queries.}

\begin{figure*}[t]
    \centering
    \includegraphics[width=\textwidth]{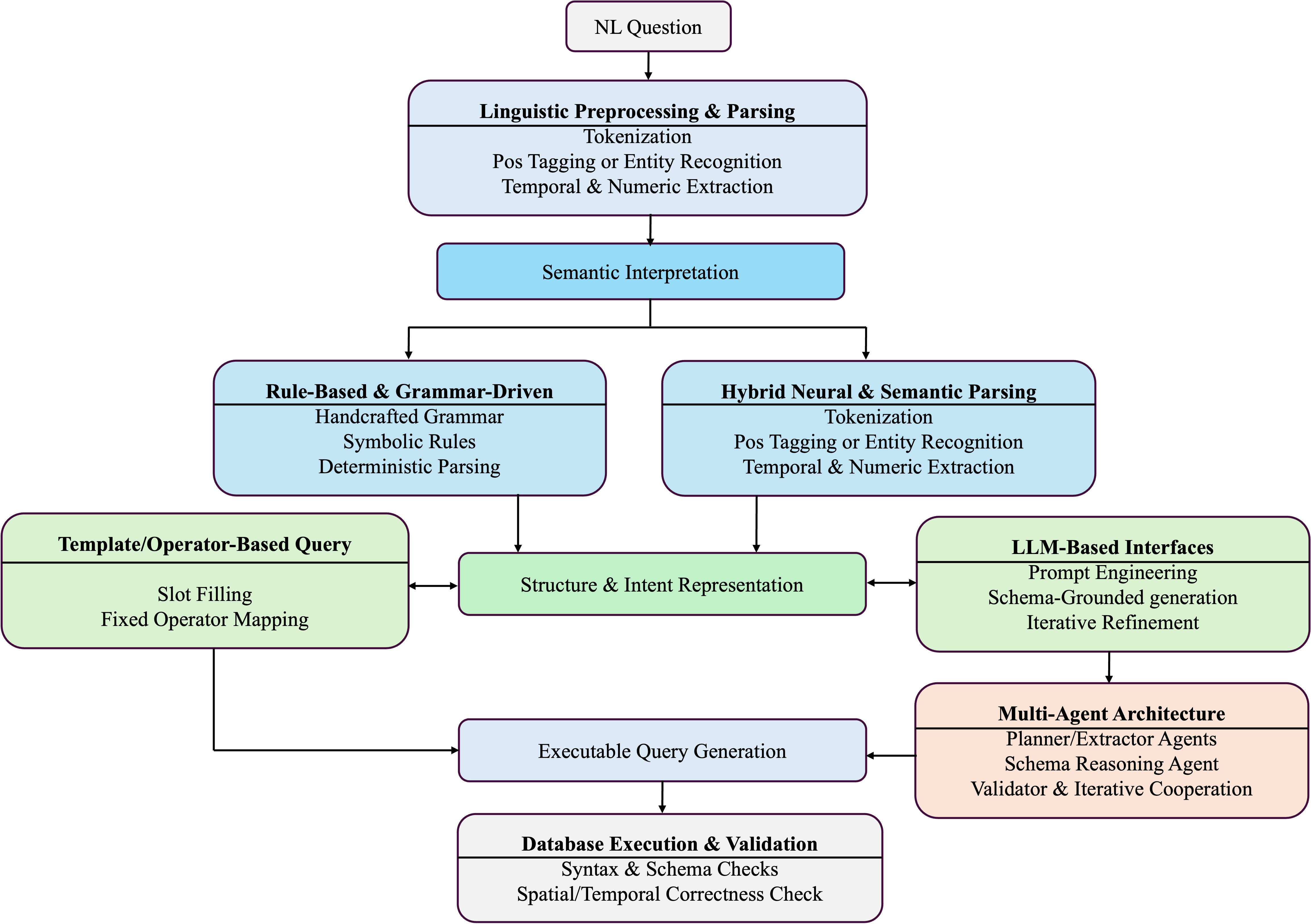}
    \caption{Unified methodological pipeline for NL interfaces to spatial, spatiotemporal, and time-series databases}
    \label{fig:method_pipeline}
\end{figure*}

\begin{table*}[t]
\centering
\small
\setlength{\tabcolsep}{5pt}
\renewcommand{\arraystretch}{1.12}
\caption{Pipeline characteristics across spatial, spatiotemporal, and time-series NLIDB systems.}
\label{tab:pipeline_by_data_type}

\begin{tabularx}{\textwidth}{@{}
p{0.15\textwidth}
p{0.26\textwidth}
p{0.26\textwidth}
p{0.26\textwidth}
@{}}
\toprule
\textbf{Pipeline Stage} &
\textbf{Spatial NLIDB} &
\textbf{Spatiotemporal / Moving Objects NLIDB} &
\textbf{Time-Series NLIDB} \\
\midrule

Natural Language Parsing &
Entity extraction focused on places, spatial relations, and numeric constraints &
Entity extraction extended with temporal expressions, object identities, and trajectory references &
Text understanding aligned to signal patterns, trends, or comparative descriptions \\

Domain Mapping &
Mapping to spatial schemas, layers, and geometry types (point, line, polygon) &
Mapping to spatiotemporal schemas, trajectory identifiers, regions, and temporal indexes &
Alignment between text descriptions and signal representations in a shared embedding space \\

Query / Task Identification &
Classification into spatial query types (range, NN, spatial join, aggregation) &
Richer classification including interval, trajectory similarity, cross-region, and detour queries &
Implicit task formulation as retrieval or comparison rather than explicit query typing \\

Query Construction Strategy &
Template-based spatial SQL or algebraic operators &
Structured spatiotemporal operators with time-aware constraints and index selection &
No explicit SQL generation; retrieval via embedding similarity and contrastive learning \\

Execution Model &
Single-shot spatial query execution &
Execution with temporal filtering, trajectory operators, and optimization-aware planning &
Similarity-based retrieval over learned representations, not database execution \\

Result Validation &
Execution success and spatial result correctness &
Execution accuracy combined with efficiency and trajectory correctness &
Retrieval accuracy based on embedding similarity and ranking metrics \\

Output Form &
Spatial objects or map-based visualizations &
Trajectories, time-filtered spatial objects, or aggregated movement summaries &
Retrieved time-series segments or ranked signal comparisons \\

\bottomrule
\end{tabularx}
\end{table*}

To better understand how ST-NLIDB systems differ across data modalities, we summarize their end-to-end processing pipelines in Table~\ref{tab:pipeline_by_data_type}. The table compares spatial, spatiotemporal, and time-series NLIDB systems by breaking them down into a common set of pipeline stages, ranging from NL parsing to final output generation.

Each row in the table corresponds to a distinct stage in the processing pipeline, describing what the system does at that step, while each column shows how this stage is executed for a particular data type. 
Spatial and spatiotemporal NLIDBs typically follow a schema-driven pipeline that culminates in executable queries, whereas time-series NLIDBs increasingly rely on representation learning and similarity-based retrieval rather than explicit query generation.

\hl{By comparing pipeline stages side by side, the table highlights both the shared structure and key differences among spatial, spatiotemporal, and time-series NLIDB systems, helping explain their differing modeling, execution, and evaluation strategies.}

The first stage typically involves NL preprocessing\cite{chowdhary2020natural} and entity extraction\cite{al2020named}. Most systems use standard NLP techniques such as tokenization, part-of-speech tagging, and named-entity recognition to identify key elements of the query, including spatial entities, temporal expressions, numeric constraints, and relational terms. Earlier and more spatially focused systems\cite{kumar2025natural}, such as studies by Xu et al.~\cite{xu2023grammar}, Zhang et al.~\cite{zhang2009natural}, and Liu et al.~\cite{liu2025nalspatial}, largely rely on rule-based extractors and manually curated vocabularies. In contrast, more recent studies increasingly incorporate neural models or LLM-based parsers to better capture implicit intent and handle more complex or flexible query phrasing, as seen in Yu et al.~\cite{yu2025monkuu} and Redd et al.~\cite{ito2024clasp}.

A second step that appears across many systems is retrieving domain information through the use of external knowledge sources. In spatial and spatiotemporal NLIDBs, this means relying on domain knowledge such as spatial relation knowledge bases, location dictionaries, database schema metadata, or prefix-tree indexes of place names. 


For spatiotemporal databases, this process is further extended to include temporal representations, trajectory identifiers, and predefined region definitions, as discussed by Wang et al.~\cite{wang2021nalmo}.

The third methodological component involves identifying the type of query. Many systems begin by classifying an NL query into one of a small set of supported categories, such as range queries, nearest-neighbor searches, spatial joins, trajectory similarity queries, or temporal interval queries. Earlier approaches typically rely on rule-based heuristics for this step, whereas more recent systems often use lightweight neural classifiers, such as LSTM-based, trained on curated collections of NL queries to infer the intended query type (Wang et al.~\cite{wang2023nalmo}, Liu et al.~\cite{liu2023nalsd}). 

After the system identifies the user’s intent, the next step is to build a structured query. In most spatial and spatiotemporal NLIDB systems, this is done using predefined templates or operator mappings rather than fully end-to-end neural model-based generation. The typical approach is slot filling, where previously extracted elements such as spatial relations, distances, time intervals, or object identifiers are inserted into fixed query templates. These slots correspond to concrete database operators such as intersects, within, k-nearest neighbors, passes, or atperiods. Using this rule-guided construction ensures that the resulting query is both syntactically valid and executable, which is important for complex spatial and temporal operations that unconstrained neural models often struggle to generate reliably.

More recent works extend this template-based pipeline by incorporating agentic or retrieval-augmented reasoning components. In these systems, complex questions are broken down into smaller sub-tasks, relevant schema information or documentation is retrieved as needed, and candidate queries are refined iteratively. Some systems also execute intermediate queries to check results before producing a final answer (Redd et al.~\cite{ito2024clasp}; Yu et al.~\cite{yu2025monkuu}; Kazazi et al.~\cite{khosravi2025questions}). Even with the introduction of LLMs, the overall pipeline remains largely unchanged, typically consisting of entity extraction, use of knowledge-bases, planning, structured query generation, execution, and result interpretation.

Most studies also include an execution-based validation step, where generated queries are run against spatial or spatiotemporal databases to confirm correctness. The outputs usually take the form of result sets, spatiotemporal trajectories, or spatial features, and are often paired with visualizations such as maps or plots. As a result, evaluation focuses on whether the query executes correctly and returns the intended results, rather than on how closely the generated query text matches a reference query.

To summarize, the literature shows that ST-NLIDB systems consistently favor hybrid architectures that combine rule-based logic, lightweight neural models, and, increasingly, LLM-based components, rather than relying on purely end-to-end learning. 

\section{Taxonomy of ST-NLIDBs}
\label{sec:method-taxonomy}


\begin{table*}[t]
\centering
\small
\setlength{\tabcolsep}{6pt}
\renewcommand{\arraystretch}{1.18}
\caption{\hl{Method families in ST-NLIDB systems: representative techniques, strengths, and limitations}}
\label{tab:method_strengths}

\begin{tabularx}{\textwidth}{p{3.5cm} p{5.5cm} X X}
\toprule

\textbf{\hl{Method Family}} & 
\textbf{\hl{Representative Approaches}} & 
\textbf{\hl{Strengths}} & 
\textbf{\hl{Limitations}} \\
\midrule

\hl{Rule-based / Grammar-driven NLIDBs} &
\hl{Rule-based NLP pipelines} \cite{woods1973progress, popescu2003towards, li2014nalir}, 
\hl{Handcrafted grammars} \cite{wang2000fuzzy, chintaphally2007extending, xu2023grammar}, 
\hl{Template-based query generation} \cite{zhang2009natural} & 
\hl{These systems translate queries through predefined linguistic rules, making the reasoning process easier to trace and the generated queries easier to inspect.} &
\hl{These systems rely on handcrafted grammars and rules, which makes them difficult to scale and adapt to diverse or evolving query formulations.} \\

\hl{Neural / Machine-learning NLIDBs} &
\hl{Seq2Seq models} \cite{zhong2017seq2sql, xu2017sqlnet}, 
\hl{LSTM-based classifiers} \cite{wang2021nalmo, liu2023nalsd, liu2025nalspatial}, 
\hl{Neural semantic parsing} \cite{zettlemoyer2012learning, yu2018typesql, wang2020rat, scholak2021picard}
&
\hl{These systems learn relationships between natural-language queries and database operations from training data, allowing them to handle linguistic variation more effectively than purely rule-based systems.} &
\hl{These systems depend on labeled training data and learned representations, which can reduce interpretability and limit generalization to unseen query types or database schemas.} \\

\hl{LLM-based NLIDBs} &

\hl{Prompt-based text-to-SQL systems} \cite{jiang2024chatgpt, ding2023gpt, pourreza2023din}, 
\hl{Schema-aware LLM prompting} \cite{wang2025gpt, yu2025monkuu, li2023resdsql, sun2023sql}
 &
\hl{These systems use LLMs to interpret natural-language questions and generate database queries through prompt-based reasoning, allowing them to handle a wide range of query expressions without relying on handcrafted rules.} &
\hl{These systems still suffer from spatial reasoning errors and may hallucinate database tables, columns, or spatial functions that do not exist in the database schema.} \\

\hl{Multi-agent NLIDBs} &

\hl{Agentic LLM pipelines} \cite{redd2025queries},
\hl{Task decomposition frameworks} \cite{khosravi2025questions, feng2025towards},
\hl{Reasoning + tool use} \cite{shelkeapproach, wang2025mac, luo2025natural, heidari2025agentiql}
&
\hl{These systems split the query task across multiple agents, allowing complex queries to be solved through step-by-step reasoning, tool use, and verification.} &
\hl{Because these systems rely on multiple interacting agents, they introduce additional system complexity, increase query latency due to multiple model calls, and require higher computational cost.} \\

\bottomrule
\end{tabularx}
\end{table*}

The ST-NLIDB systems differ in how they translate natural language queries into executable spatial and temporal database queries. Across all ST-NLIDB systems, we can see four main methodological families: (i) rule-based and grammar-driven systems, (ii) neural net and semantic parsing-based hybrid systems, (iii) LLM prompting approaches, and (iv) multi-agent architectures. These methodological differences show a gradual shift from more determinisitic, hand-engineered pipelines toward flexible and agentic frameworks. This section details the methodology used by each category of NLIDBs. For each category, we begin by giving an overview of how general-purpose NLIDBs translate NL questions into executable queries, and then we examine how ST-NLIDBs extend the corresponding approach to handle spatial and temporal semantics.

\hl{Each methodological family offers distinct advantages and limitations in translating natural language questions into executable spatial and temporal database queries. These differences reflect the evolution of ST-NLIDB systems from early rule-based pipelines to learning-based and agentic approaches. Table} ~{\ref{tab:method_strengths} \hl{summarizes these four methodological categories, along with their representative techniques, main strengths, and limitations.}

\subsection{Rule-Based and Grammar-Driven NLIDBs}
\label{subsec:rule-grammer-driven}
In rule/grammar-based systems, the interpretation of a user’s question is governed by explicit linguistic rules or formal grammars that describe how valid spatial queries can be expressed. Instead of learning mappings from data, the system relies on manually defined grammar productions, semantic rules, and domain knowledge to identify entities, spatial relations, and query intent. Once the input sentence matches a known grammatical structure, it is deterministically translated into a structured spatial query or analytical workflow. These approaches mainly focus on the syntactic structure of the NL question and the database query. In non-spatiotemporal general-purpose NLIDBs~\cite{woods1973progress, lin2019grammar, yu2025spatial, deng2025reforce, wang2025mac, zhai2025optimizing, ghosh2025sqlgenie, feng2025engineering, patel2025natural, sharma2025classifying, li2014nalir, popescu2003towards} under this category, an NL query is parsed based on its syntactic structure, and a parse (syntax) tree is generated. The parse tree nodes are mapped to database nodes for the purpose of generating a database query. The syntax structure captured by the parse tree plays a significant role in generating the corresponding database counterpart for the NL query. While this approach offers transparency and control over query interpretation, it also requires substantial manual effort to design and maintain the grammars, and it typically struggles to handle linguistic variation or unexpected query formulations.

Rule-based and grammar-driven natural language interfaces represent some of the earliest attempts in spatial NLIDB research. Early research on NL interfaces for spatial databases relied heavily on rule-based reasoning and grammar-driven parsing, where the structure of a user query was explicitly defined through linguistic rules. These approaches were motivated by the limited availability of large training datasets and the need for predictable, interpretable behavior in spatial query translation.

One of the earliest examples is the work by Wang et al. (2000)\cite{wang2000fuzzy}, where a fuzzy grammar and possibility theory–based NL interface were proposed for spatial queries. The system focused on handling the inherent vagueness of natural language, such as terms like near or far, by combining fuzzy grammar rules with possibility theory. The input sentence was first parsed using a predefined fuzzy grammar, after which vague spatial expressions were converted into quantitative representations using fuzzy logic. The final interpretation was then translated into an SQL query executable on a spatial database. This work highlighted the importance of explicitly modeling linguistic uncertainty in spatial queries, although the approach required substantial manual rule design.

Following this, Chintaphally et al. (2007)\cite{chintaphally2007extending} extended traditional NL interfaces by incorporating geospatial querying capabilities. Their system employed a menu-based, predictive interface constrained by a domain-specific attributed grammar. Users were guided to construct valid NL queries ensuring grammatical correctness. The grammar rules directly mapped NL phrases to relational joins and spatial joins, which were then translated into Oracle Spatial SQL query. Query results were returned in Geography Markup Language (GML) format and further converted to Keyhole Markup Language (KML) for visualization in tools such as Google Earth Engine. While this approach reduced parsing ambiguity, it also limited user expressiveness by restricting queries to predefined grammatical patterns.

In a similar rule-driven work, Zhang et al. (2009)\cite{zhang2009natural} presented a natural language interface for crime-related spatial queries. Their method combined part-of-speech tagging and syntactic parsing with rule-based semantic interpretation. Each query was decomposed into a semantic triplet consisting of a target object, a spatial predicate, and a reference object. These components were then matched to database schemas using weighted string similarity, followed by spatial predicate evaluation within the database. The results were visualized using KML. This work demonstrated how carefully designed semantic rules could support real-world, domain-specific spatial querying, although the system depended strongly on handcrafted rules and domain assumptions.

More recently, Xu et al. (2022)\cite{xu2023grammar} revisited grammar-based methods from a more conceptual perspective in a grammar for interpreting geo-analytical questions as concept transformations. Rather than directly generating SQL, their approach aimed to capture the analytical intent behind geo-analytical questions. Natural-language input was processed using named entity recognition and mapped to core spatial concepts such as objects, fields, and events. A context-free functional grammar was then used to identify functional roles within the question, producing a parse tree that was transformed into a directed acyclic graph of concept transformations. The output was an abstract GIS workflow specification, independent of any specific GIS platform. This work emphasized interpretability and analytical clarity, but still relied on manually defined grammars and concept dictionaries.

Overall, rule-based and grammar-driven approaches laid the foundation for natural language interaction with spatial databases. They offered transparency and deterministic behavior but required extensive manual effort to design grammars, rules, and domain knowledge. As spatial queries grew more complex and diverse, these limitations motivated later shifts toward learning-based and LLM-driven methods.

Figure~\ref{fig:grammar_flow} shows a flow diagram that summarizes these methods. It captures the common steps that appear across those papers, even though each system implements them in its own slightly different way.

\begin{figure}[!t]
    \centering
    \includegraphics[width=\linewidth]{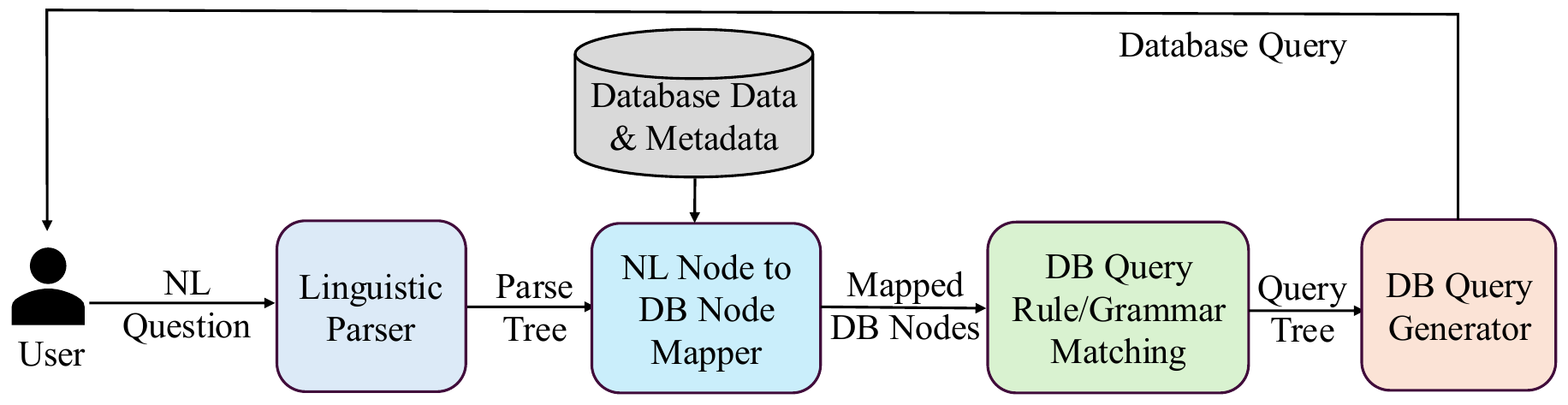}
    \caption{General flow structure of rule-based and grammar-driven NLIDBs}
    \label{fig:grammar_flow}
    \vspace{-16pt}
\end{figure}

\vspace{4pt}

\vspace{5pt}
\noindent\fbox{%
  \begin{minipage}{\columnwidth}
   
\textbf{Key Takeaway}: \textit{\hl{Rule-based and grammar-driven NLIDBs represent early approaches to natural-language querying of spatial databases, where queries are interpreted using manually defined linguistic rules and grammars. While this design provides clear and predictable query translation, the reliance on handcrafted rules makes these systems difficult to scale as query language and database schemas become more complex.}}

  \end{minipage}%
}

\subsection{Neural Net and Semantic Parsing-Based Hybrid NLIDBs}
As natural language interfaces for databases evolved, researchers began to move beyond fully rule-based systems toward neural and semantic parsing-based hybrid models. The idea of these approaches were initially introduced by generic non-spatiotemporal NLIDBs~\cite{zettlemoyer2012learning, kim2010generative, kate2007learning, zhong2017seq2sql, xu2017sqlnet, yu2018typesql, yu2018syntaxsqlnet, wang2020rat, scholak2021picard}. These approaches still preserve a structured view of spatial querying but introduce machine learning components to handle linguistic variability, ambiguity, and scalability. Instead of relying solely on handcrafted grammars, hybrid systems typically combine traditional NLP preprocessing and domain knowledge with neural models, most commonly for tasks such as entity extraction, query type classification, or semantic disambiguation. However, the final query generation step, often remains template-driven or operator-based, ensuring execution reliability while benefiting from learned representations.

An early example of this transition in the moving objects domain is NLMO by Xu et al. (2020)\cite{wang2020nlmo}. This system applied standard NLP preprocessing and named-entity recognition to extract temporal expressions, numeric values, and object identifiers from natural-language queries. A key contribution was the use of a learned classifier to distinguish between different spatiotemporal query types, such as range and nearest-neighbor queries. Based on the predicted type, the system mapped extracted entities into predefined spatiotemporal operators and generated executable queries for the SECONDO moving-objects database. While the overall translation process remained structured, the introduction of learning-based components improved robustness compared to purely rule-based systems.

Building on similar ideas, Wang et al. (2021)\cite{wang2021nalmo} proposed NALMO, a natural language interface specifically designed for moving object databases. Their approach combined spaCy-based semantic parsing and entity recognition with a location knowledge base to resolve spatial references. An LSTM-based classifier was trained to identify query intent categories, such as time interval, range, or trajectory similarity queries. Once the query type was identified, extracted entities were mapped to corresponding SECONDO operators using predefined rules. This hybrid design allowed the system to balance interpretability and flexibility while maintaining reliable execution on spatiotemporal data.

In a later extension, Wang et al. (2023)\cite{wang2023nalmo} further refined the NALMO framework by improving entity normalization and semantic parsing. The system incorporated more detailed handling of temporal expressions, numeric normalization, and relation extraction using domain rules, followed by LSTM-based query type prediction. Compared to earlier versions, this work focused on better alignment between natural language expressions and structured spatiotemporal operators, particularly for more complex trajectory-based queries. Despite these improvements, the system still relied on predefined operator compositions, reflecting the structured nature of hybrid semantic parsing approaches.

Parallel developments occurred in the spatial database domain. Liu et al. (2023)\cite{liu2023nalsd} introduced NALSD, a natural language interface that combined neural query classification with template-based spatial query generation. Their system first performed tokenization and named-entity recognition to extract spatial relations, locations, and numeric constraints. An LSTM classifier then determined whether the query corresponded to a range query, nearest-neighbor query, or spatial join. Based on this classification, the system selected appropriate spatial operators and generated executable SECONDO queries. This design demonstrated how limited neural components could significantly reduce the manual burden of rule specification while retaining structured execution.

In the same year, Liu et al. (2023)\cite{liu2023nalspatial} also proposed NALSpatial, which followed a similar hybrid architecture but placed greater emphasis on knowledge-base support. The system leveraged both a location knowledge-base and a spatial-relation knowledge-base to refine entity extraction and disambiguation. A manually curated corpus of spatial NL questions was used to train an LSTM classifier for query type identification. Once classified, queries were translated using predefined templates with placeholders for spatial relations, locations, distance thresholds, and aggregation parameters. This work highlighted the effectiveness of combining learned intent classification with explicit domain knowledge in spatial query translation.

More recently, Liu et al. (2025)\cite{liu2025nalspatial} extended the NALSpatial framework with a distinct separation between natural language understanding and translation phases. Coarse entity extraction was followed by refinement using knowledge bases, after which an LSTM classifier predicted the query category. A structured language model then mapped extracted entities into operator templates corresponding to different spatial query types. While neural components played an important role in interpretation, the execution logic remained deterministic, reflecting the hybrid nature of the approach.

Finally, NALMOBench by Wang et al. (2025)\cite{wang2150nalmobench} provided both a benchmark and a strong baseline representative of this category. The baseline system employed a two-stage entity extraction process, followed by neural query type classification and structured slot filling. Although this study mainly proposed an evaluation framework, the baseline architecture closely mirrors earlier hybrid semantic parsing systems, reinforcing this design pattern as a dominant approach prior to the widespread adoption of large language models.

Overall, neural and semantic parsing-based hybrid models lie in the middle of early grammar-based systems and modern LLM-driven approaches. By combining learned components with structured query generation, these systems improved robustness and scalability while preserving control over spatial query execution. However, their reliance on predefined query types and operator templates still limits expressiveness, motivating later shifts toward end-to-end and agent-based LLM solutions.

\hl{Figure}~{\ref{fig:nn_model} \hl{illustrates the general pipeline used by neural net and semantic parsing–based hybrid NLIDBs, where natural language queries are processed through NLP preprocessing, neural classification, and template-based query generation.}

\begin{figure}[!t]
    \centering
    \includegraphics[width=\linewidth]{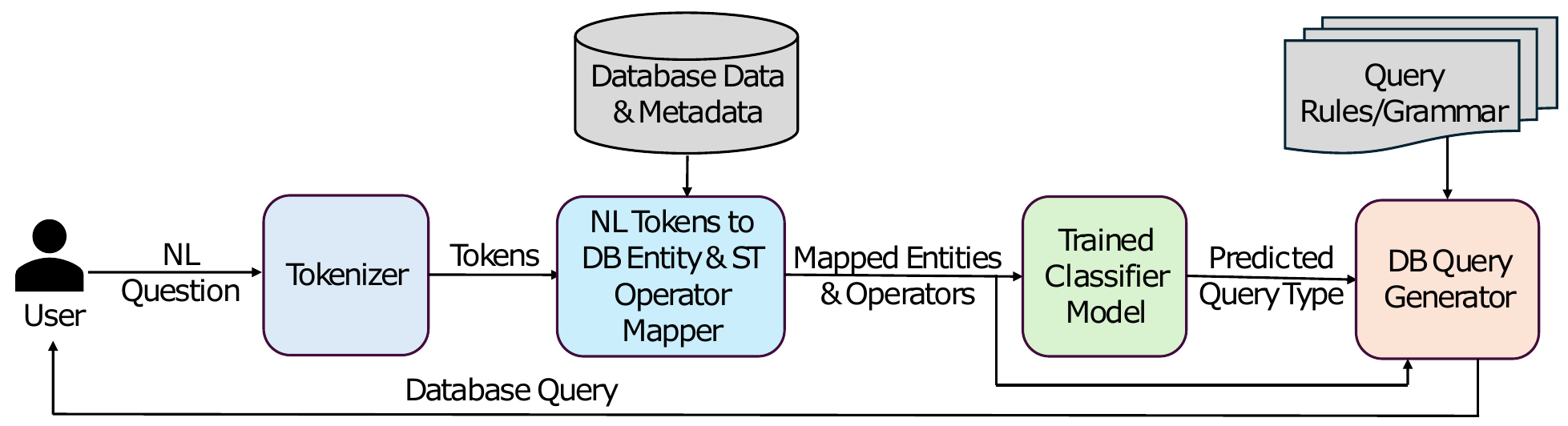}
    \caption{Generic pipeline of neural net and semantic parsing-based hybrid NLIDBs}
    \label{fig:nn_model}
\end{figure}


\vspace{5pt}
\noindent\fbox{%
  \begin{minipage}{\columnwidth}
   
\textbf{Key Takeaway}: \textit{\hl{Neural net and semantic parsing–based hybrid NLIDBs combine machine-learning models with structured query templates to interpret natural-language questions. Learning components help identify entities and query types, while template-based translation keeps query execution reliable. However, because these systems still rely on predefined templates and operators, they struggle with spatial queries that fall outside the expected patterns.
}}

  \end{minipage}%
}

\subsection{LLM-Based NLIDBs}
With the emergence of large language models, research on natural language interfaces for databases began to shift away from rigid templates and task-specific neural models toward LLM-driven interpretation and generation. In these approaches (non-spatiotemporal)~\cite{hazoom2021text, fan2024combining, ding2023gpt, pourreza2023din, sun2023sql, li2023resdsql, pham2025multilingual}, the LLM plays a central role in understanding user intent, relating NL queries in database schemas, and generating executable spatial queries. Unlike earlier hybrid systems, LLM-based interfaces rely primarily on prompt design, in-context learning, and semantic reasoning rather than explicit query type classification or handcrafted operator mappings.

In the spatial and temporal domain, one of the earliest studies exploring this direction is Jiang et al. (2023)\cite{jiang2024chatgpt}, who investigated whether ChatGPT can function as a geospatial data analyst. Their system treated spatial query translation as a prompt-based text-to-SQL task. Database schemas and sample data were provided to the LLM using CREATE TABLE statements and example rows, along with the user’s NL question. The LLM then generated PostGIS SQL queries directly, which were executed on a spatial database. Query results were subsequently transformed into natural-language responses. This work demonstrated the feasibility of using general-purpose LLMs for spatial SQL generation, while also highlighting issues related to execution reliability.

Building on this idea, Wang et al. (2025)\cite{wang2025gpt} proposed a more structured prompt-engineering strategy in GPT-Based Text-to-SQL for Spatial Databases. Their approach decomposed the prompt into multiple components, including spatial database knowledge, in-context examples, schema definitions, geographic descriptions, and feedback derived from previous execution errors. If the generated SQL failed during execution, schema-related errors were extracted and injected back into the prompt as guidance for subsequent iterations. This iterative prompting mechanism improved execution success rates compared to single-shot generation, while still maintaining a single-LLM pipeline.

More advanced schema handling was introduced by Yu et al. (2025)\cite{yu2025monkuu} in Monkuu, which focused on dynamic schema mapping for geospatial databases. Instead of directly exposing full schemas to the LLM, Monkuu generated compact schema abstractions from expert-verified documentation and used the LLM to map query semantics to relevant tables and columns. The system also incorporated human-in-the-loop disambiguation for geographic entities when ambiguity was detected. Once entities and schema elements were resolved, the GIS-aware SQL were generated with LLM using few-shot prompting with spatial function guidance. This work addressed one of the major limitations of earlier LLM-based systems, namely poor scalability to complex or evolving schemas.

In parallel, Yu et al. (2025)\cite{yu2025spatial} also proposed Spatial-RAG, which extended LLM-based interfaces beyond direct SQL generation. Instead of relying solely on query translation, Spatial-RAG combined spatial database retrieval with embedding-based semantic scoring. The system first used an LLM to parse spatial intent and generate a spatial SQL query to retrieve candidate objects. These candidates were then ranked using both spatial relevance metrics and semantic similarity derived from embeddings. A Pareto-based filtering step was applied to balance spatial and semantic relevance, after which the LLM selected the final answer and generated a natural-language response. This approach illustrated how LLMs can be integrated with retrieval-augmented reasoning rather than acting purely as SQL generators.

A related but domain-specific application of LLM-based interfaces was presented by Sezgin et al. (2025)\cite{wang2025towards} in the context of Internet of Drones (IoD) platforms. Their system used sentence-level embeddings and vector search to retrieve relevant API examples, which were then incorporated into a few-shot prompt for an LLM. The LLM inferred the appropriate API endpoint and parameters, generating structured JSON calls validated against OpenAPI specifications. Although not strictly focused on spatial SQL, this work demonstrated how LLM-based interfaces can support structured query generation over spatially systems with strong schema and safety constraints.

Overall, LLM-based natural language interfaces represent a significant shift from earlier semantic parsing paradigms. By leveraging pretrained language models and prompt-based reasoning, these systems reduce the need for task-specific training data and handcrafted rules. \hl{LLM-based NLIDBs provide a high level of flexibility, but they also come with several limitations due to the nature of generative models. One key challenge is geo-hallucination} {\cite{huang2025geo}, \hl{a form of spatial reasoning error that results in the generation of spatial queries with incorrect location mappings.} \hl{Beyond geo-hallucination, prompt brittleness} {\cite{ngweta2025towards} \hl{is another challenge. Small changes in how a user describes a geographic region or a time range can lead the model to produce very different query plans. Another practical limitation is computational cost. LLM-based NLIDBs often rely on long prompts that include database schemas, spatial metadata, and example queries. These large prompts increase inference latency and computational cost, particularly when external LLM services are used for real-time query generation. Privacy is another important concern. Since schema details or example data are sometimes included in prompts, sensitive information about the database structure or its underlying data could be exposed, especially when external LLM services are used} {\cite{liu2025safenlidb}.
As a result, challenges remained in execution reliability and control, which have motivated further developments toward multi-agent architectures.

\begin{figure}[!t]
    \centering
    \includegraphics[width=\linewidth]{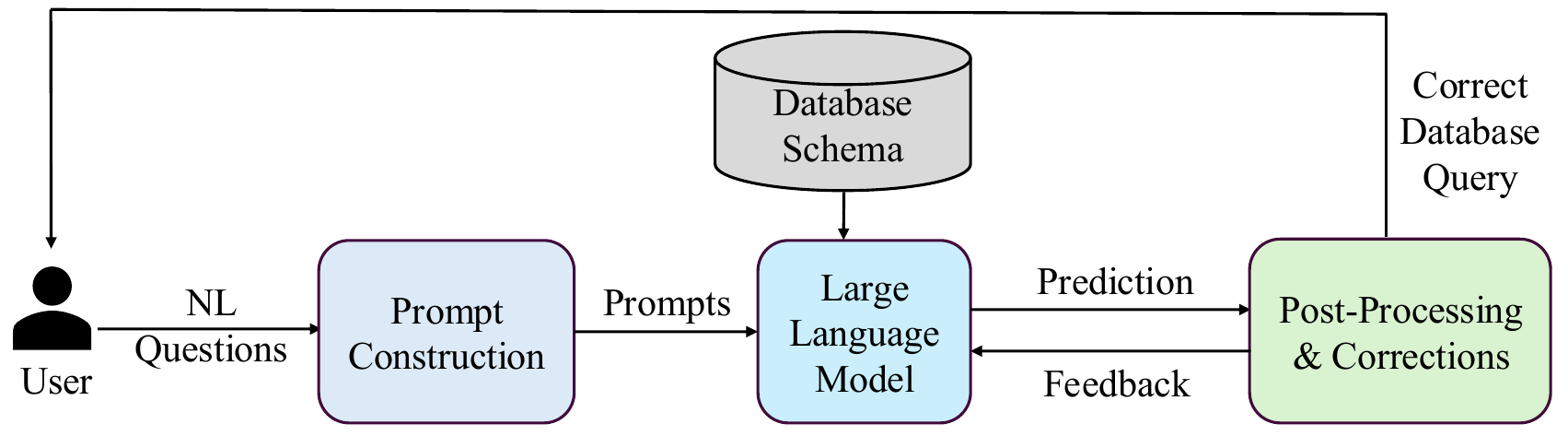}
    \caption{A general illustration of how LLM-prompting approaches generate database queries.}
    \label{fig:llm_prompt}
\end{figure}

\hl{Figure}~{\ref{fig:llm_prompt} \hl{shows the general workflow of LLM-based NLIDBs, where database schemas and example queries are included in the prompt to guide the generation of executable spatial SQL queries.}


\vspace{5pt}
\noindent\fbox{%
  \begin{minipage}{\columnwidth}
   
\textbf{Key Takeaway}: \textit{\hl{LLM-based NLIDBs use large language models to interpret natural-language questions and generate database queries through prompt-based reasoning. This allows them to handle a wide variety of query expressions without relying on handcrafted rules or predefined templates. However, due to geo-hallucination and prompt brittleness, the system may sometimes produce incorrect SQL or reference tables, columns, or spatial functions that do not exist in the database schema.
}}

  \end{minipage}%
}

\subsection{Multi-Agentic NLIDBs}
More recent work has moved beyond single-model pipelines toward multi-agent architectures, where the task of interpreting and executing spatial or spatiotemporal queries is explicitly decomposed across multiple cooperating components. In these systems, different agents are assigned specialized roles such as entity extraction, planning, schema reasoning, query generation, validation, and explanation. Rather than relying on a single pass of query generation, multi-agent approaches emphasize iterative reasoning, verification, and correction, aiming to improve robustness, interpretability, and execution reliability for complex queries. In the non-spatiotemporal domain, multi-agentic approach was adopted by NLIDBs~\cite{shelkeapproach, wang2025mac, luo2025natural, rauch2025conversational, li2025sql, heidari2025agentiql}.

\begin{figure}[!t]
    \centering
    \includegraphics[width=\linewidth]{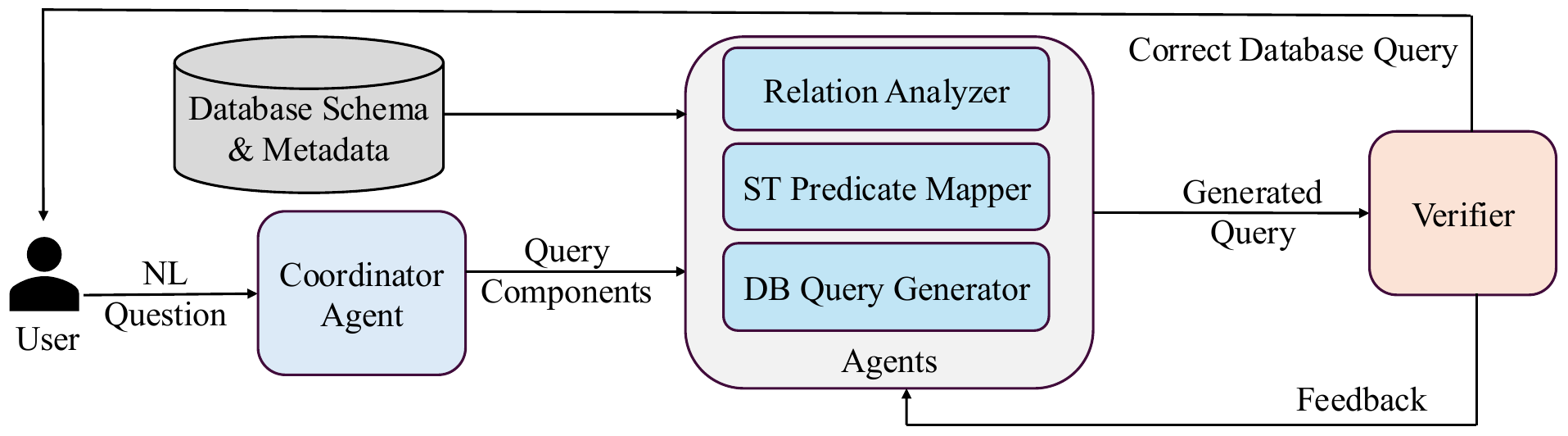}
    \caption{A general illustration of how multi-agent approaches generate database queries.}
    \label{fig:multi_agent}
\end{figure}

\hl{Figure}~{\ref{fig:multi_agent} \hl{illustrates the general workflow of multi-agentic NLIDBs, where different agents cooperate to interpret the query, generate SQL statements, and validate the final results.}

In the spatiotemporal domain, an early example of this direction is the work by Redd et al. (2025)\cite{redd2025queries}, who proposed an agentic LLM pipeline for spatiotemporal text-to-SQL. In this system, an LLM agent alternates between reasoning, action, and observation. Given an NL query, the agent first interprets the intent and inspects the database schema, including tables, columns, and sample rows. A baseline text-to-SQL model is then invoked as a tool to generate an initial query, which is executed and evaluated. If errors occur or results are empty, the agent refines the query through repeated plan–act–observe cycles, potentially decomposing the task or injecting external knowledge such as spatial boundaries or temporal constraints. This approach highlights how agentic control can compensate for the brittleness of single-shot LLM query generation.

A more explicitly structured multi-agent design was presented by Kazazia et al. (2025)\cite{khosravi2025questions}, where the framework assigns distinct responsibilities to separate agents, including entity extraction, metadata retrieval, query logic construction, SQL generation, and review. The system first validates the user query and coordinates agent execution through an orchestration component. Extracted entities are mapped to database schemas using metadata embeddings, after which a query logic agent constructs an abstract logical plan describing joins, spatial predicates, and aggregations. This plan is translated into PostGIS SQL by a dedicated generation agent and then verified by a review agent that checks syntax, schema consistency, and spatial function correctness. Errors trigger automatic regeneration, resulting in a validated spatial SQL query. This work demonstrates a separation of concerns within the pipeline of query translation system, improving control and debuggability.

In a broader GeoQA context, Feng et al. (2025)\cite{feng2025towards} proposed a multi-agent architecture aimed at building a barrier-free geospatial question-answering portal. Their system begins with a router agent that classifies user intent and dispatches the query to analysis, explanation, or visualization modules. Within the analysis path, specialized agents perform semantic parsing, task decomposition, region selection through geocoding, and hybrid entity retrieval using keyword matching, vector search, and LLM-guided disambiguation. Spatial operations are executed using predefined geometry processing functions, and intermediate results are stored for reuse. Separate explainer and visualizer agents generate textual explanations and interactive map outputs, respectively. This design emphasizes user interaction and transparency, allowing users to inspect intermediate steps rather than treating the system as a black box.

Overall, multi-agent natural language interfaces represent a shift toward explicit reasoning and control in spatial query processing. By decomposing complex tasks into coordinated subtasks and incorporating verification loops, these systems address many of the limitations observed in earlier LLM-based pipelines, particularly in terms of reliability and error handling. However, the added architectural complexity and computational cost introduce new challenges, suggesting a trade-off between robustness and system simplicity.

\vspace{5pt}
\noindent\fbox{%
  \begin{minipage}{\columnwidth}
   
\textbf{Key Takeaway}: \textit{\hl{Multi-agentic NLIDBs approach the query translation problem by dividing it into several smaller tasks handled by different agents, such as understanding the query, reasoning over the schema, generating the SQL, and checking the result. This collaborative process helps the system handle more complex spatial queries and improves reliability. However, coordinating multiple agents also makes the system more complex and usually requires several model calls to complete a single query.}}

  \end{minipage}%
}

\begin{table*}[t]
\centering
\small
\setlength{\tabcolsep}{3.8pt}
\renewcommand{\arraystretch}{1.08}
\caption{Methodological characteristics of natural language interfaces for spatial, spatiotemporal, and time-series databases.}
\label{tab:method_taxonomy}

\begin{tabularx}{\textwidth}{@{}
p{0.16\textwidth}
*{6}{>{\RaggedRight\arraybackslash}X}
>{\centering\arraybackslash}p{0.07\textwidth}
@{}}
\toprule
\textbf{Method Class} &
\textbf{Interpretation Mechanism} &
\textbf{Learned Components} &
\textbf{Schema Understanding} &
\textbf{Query Planning} &
\textbf{Execution \& Validation} &
\textbf{Ambiguity Handling} &
\textbf{Extensibility} \\
\midrule

Rule-Based \& Grammar-Driven &
Handcrafted grammars and rules &
None &
Manual, rule-based mapping &
Deterministic &
Built-in constraints &
Explicit rule encoding &
Low \\

Neural \& Hybrid Semantic Parsing &
Neural classifiers with structured templates &
Partial (e.g., LSTM, embeddings) &
Knowledge-base assisted &
Template-driven &
Deterministic execution &
Learned + rule-based &
Medium \\

LLM-Based Interfaces &
Prompt-based language reasoning &
Fully learned &
Prompt- and schema-context understanding &
Generative &
Limited or iterative &
Implicit &
High \\

Multi-Agent Architectures &
Coordinated agent reasoning &
Fully learned &
Agent-mediated schema reasoning &
Planned and iterative &
Explicit agent validation &
Explicit reasoning steps &
Very High \\
\bottomrule
\end{tabularx}
\end{table*}

\subsection{Summary of ST-NLIDBs Taxonomy}
To provide a structured overview of how natural language interfaces to spatial, spatiotemporal, and time-series databases are designed, we categorize the reviewed systems according to their underlying methodological characteristics. Rather than focusing on individual implementations, this taxonomy highlights the core design choices that differentiate existing approaches, including how NL queries are interpreted, how learning is incorporated and how queries are planned and executed. Table~\ref{tab:method_taxonomy} summarizes these methodological dimensions and organizes prior work into four broad method classes, ranging from traditional rule-based systems to recent large language model and multi-agent architectures. This abstraction allows common patterns and trade-offs to be identified across systems that otherwise differ substantially in scope, data modality, and technical realization.

\section{Discussion}
\label{sec:discussions}
The research in natural language interfaces for spatial, spatiotemporal, and time series databases has grown in several varying directions, but there are still limitations.
This section focuses on our observations and the limitations we noticed across the existing studies. The limitations can be seen on both the datasets and the methodological perspectives, and they often appear together in ways that affect how well these systems can be used in real settings.

\subsection{Limitations of Existing Datasets}
\label{sec:datasets-limitations}
One major limitation is the lack of a standard, widely used benchmark for this domain, unlike general NL2SQL research, which has widely adopted datasets such as WikiSQL\cite{hwang2019comprehensive} and Spider\cite{yu2018spider}. Most datasets used in ST-NLIDBs are small, synthetic, or tied to particular database environments, making it difficult to conduct fair comparisons across different systems. Recent works such as GeoSQL Bench and GeoSQL Eval \cite{houanl2sql, hou2025geosql} provide a large set of synthetic tasks and help with operator coverage, but they still do not capture real geographic irregularities or authentic user queries. Other datasets, such as SpatialQueryQA \cite{khosravi2025questions} and GeoQueryJP \cite{yu2025monkuu}, are useful for testing specific situations, but they are relatively small and lack the range of spatial and temporal operators needed for an evaluation representing real-world conditions.


Spatiotemporal datasets like BerlinMOD \cite{duntgen2009berlinmod} and the SECONDO moving-object collections used in NLMO~\cite{wang2020nlmo} and NALMO \cite{wang2021nalmo, wang2023nalmo} cover only a narrow slice of spatiotemporal reasoning, mostly focusing on movement patterns or time windows. Time-series datasets such as TRUCE, SUSHI, and the TACO-based pairs \cite{ito2024clasp, dohi2025retrieving} support natural-language supervision for temporal patterns, but they are not linked to any spatial geometry. Therefore, they cannot test how systems combine time with location. Across all these categories, noisy place descriptions and multi-step reasoning, which are common in real applications, remain mostly untested.

\hl{Another limitation is that existing ST-NLIDB datasets rarely include ambiguous or underspecified} {\cite{wang2023know,wu2024need,zhang2020did} \hl{user queries. Real-world questions often contain vague spatial or temporal expressions such as “near”, “around”, or “recent”, where the intended distance or time range is not clearly defined. However, such evaluation settings are still largely missing in ST-NLIDB benchmarks.} \hl{Finally, current ST-NLIDB benchmarks do not test systems under schema anonymization or obfuscation} {\cite{fieschi2025characterising,templ2022systematic} \hl{. To handle sensitive information, ST-NLIDB benchmarks can replace table and column names with neutral identifiers. However, this type of evaluation is still missing in ST-NLIDB benchmarks.}

Putting everything together, current ST-NLIDBs lack a common evaluation and benchmarking system. Many systems perform well on the datasets they were trained or designed around, but their generalization to new cities or more natural user queries remains uncertain. The absence of a large, widely adopted benchmark that covers spatial and/or temporal reasoning is therefore one of the central limitations of current work.

\subsection{Limitations of Current Methods}

\subsubsection{Rule-based and grammar systems break easily with schema or language changes}
Rule-based and grammar-driven approaches \cite{chintaphally2007extending, zhang2009natural, xu2023grammar} are highly sensitive to phrasing and often require manual updates whenever the schema or vocabulary shifts. They do not adapt smoothly when new geographic layers are introduced or when users phrase queries in a more informal or inconsistent way.

\subsubsection{Neural net and semantic parsing-based models rely on small, domain-Specific Corpora}
Neural and Semantic systems \cite{li2019spatialnli, wang2021nalmo, wang2023nalmo} reduce some of the rigidity of rule-based designs, but they are still trained on small, highly domain-specific datasets. Many spatial reasoning steps continue to depend on handcrafted constraints, which limits scalability and makes the models struggle with varied geographic structures or unseen spatial operators.

\subsubsection{LLM prompting lacks deep spatial understanding and is highly prompt-sensitive}
LLM prompting methods \cite{jiang2024chatgpt, wang2025gpt} may choose spatial operators based on surface-level word similarity rather than real geometric understanding, and hallucination remains a challenge \cite{ji2023towards}. They also depend heavily on schema descriptions and carefully engineered prompts, and the quality of the output shifts significantly depending on how the prompts are phrased.

\subsubsection{Multi-agent pipelines are powerful but prone to coordination errors and high cost}
Multi-agent systems \cite{khosravi2025questions, feng2025towards, redd2025queries} introduce heavier computational cost and require careful coordination between model agents and the environment. If any agent misinterprets schema information or generates an incorrect intermediate reasoning step, the entire pipeline can drift away from the intended query logic. Their performance depends on the capabilities of the underlying LLM and on how well each agent’s role is defined.

\subsubsection{ST-NLIDBs lack adoption of geospatial representation learning techniques}
Recent NL2SQL research \cite{luo2025natural, zhan2025natural} in general relational databases now uses advanced neural ideas such as schema linking through learned embeddings \cite{kaygudetext}, cross-attention between natural language and tables \cite{liu2025graph}, and large-scale SQL pretraining. Systems such as RAT-SQL \cite{lyu2025sql}, LGESQL \cite{zhao2025enhancing}, T5-based text-to-SQL architectures \cite{singh2025survey}, dense retrieval and representation learning approaches \cite{li2025detriever}, and broader NL2SQL models \cite{srikanth2025sqlspace} have been adopted widely. None of the spatial systems incorporate similar techniques, even though spatial data naturally contains strong geometric and topological structure that could benefit from learned embeddings.

\textbf{Overall Gap}: Existing methods still lack full spatial and temporal reasoning capabilities. Across all these categories, there is still a gap between the complexity of spatiotemporal reasoning and what current systems provide.
Representation learning techniques widely adopted in NL2SQL research have not yet been integrated into ST-NLIDBs, leaving room for significant improvement.

\subsection{Future Research Directions}
Our findings suggest that research on ST-NLIDBs have made good progress, but there are still room for major improvements. The main gap between controlled experiments and real-world use means that future systems will need to focus more on robustness rather than only accuracy on a small benchmark.

One important area for future work is the development of stronger and more consistent benchmark datasets, similar to WikiSQL\cite{hwang2019comprehensive} and Spider\cite{yu2018spider} for general-purpose NLIDBs. A benchmark that combines spatial and/or temporal reasoning, noisy geographic descriptions, and realistic schema variation would better represent the real-world scenarios.
\hl{Another useful direction is the design of benchmarks that include ambiguous or underspecified queries, as discussed in Section} \ref{sec:datasets-limitations}.
\hl{New benchmarks can also be designed to evaluate systems under schema anonymization or obfuscation. Such settings would help assess whether ST-NLIDB systems rely on schema names or actually interpret the query.}
In addition, geospatial representation learning techniques and advances from general NL2SQL research, such as schema linking, cross-attention between text and tables, or pretraining on SQL logs, may be integrated into ST-NLIDB systems.

Databases such as PostGIS~\cite{postgis} also support managing satellite imagery datasets. Therefore, It is time to generate NLIDB benchmarking datasets for remote-sensing images and extend the NLIDB research in this direction. 

Overall, the future work in ST-NLIDBs should focus on building systems that are not just accurate in simplified settings but are also more generalizable and able to deal with the messy and sometimes unpredictable nature of real-world geographic data.

\section{Conclusion}
This article presents a comprehensive review of natural language interfaces for spatial and temporal databases. We first outlined the representative database systems and defined the corresponding NLIDB tasks through illustrative example queries. We then enumerated the datasets and evaluation metrics in the literature of spatiotemporal NLIDBs. After defining the common methodological pipelines in NLIDBs in this domain, we organized the taxonomy and analyzed every class of methods. Finally, we discussed key limitations of current research, spanning data coverage, modeling assumptions, and practical deployment considerations, and outlined promising research directions for future work. Collectively, these contributions provide a unified perspective on the state of the art and offer guidance for advancing NLIDBs for spatial and temporal data management.

\bibliographystyle{IEEEtran}
\bibliography{reference}

\EOD
\end{document}